\documentclass[twocolumn]{aastex63}
\usepackage{amsmath,amstext}
\usepackage{mathtools}
\usepackage{afterpage}
\usepackage{lipsum}
\usepackage{capt-of}
\usepackage{relsize}
\usepackage{newtxtext,newtxmath}
\usepackage[all]{hypcap} %Links go to figures. This breaks deluxetables; use \capstartfalse \capstarttrue around deluxetables to fix it.
\usepackage{breqn}
\usepackage{verbatim}
\usepackage{lineno}
\usepackage{graphicx}
\usepackage{footnote}
\usepackage{float}
\usepackage{array,multirow}
\usepackage{amsmath}
\usepackage{enumitem}
\usepackage{amssymb}
\usepackage{array}
\usepackage{gensymb}
\usepackage{mathtools}
\usepackage{relsize}
\usepackage{lipsum}  
\usepackage[graphicx]{realboxes}
\usepackage{natbib}

\usepackage{lipsum}
\usepackage{rotating}
\usepackage{nicefrac}
\newcommand\blfootnote[1]{%
  \begingroup
  \renewcommand\thefootnote{}\footnote{#1}%
  \addtocounter{footnote}{-1}%
  \endgroup
}
\usepackage{color, colortbl}
\newcommand{\CAPS}{Center for Astrophysical Surveys, Urbana, IL, 61801, USA}

\newcommand{\Illinois}{Department of Astronomy, University of Illinois at Urbana-Champaign, 1002 W. Green St., IL 61801, USA}
\newcommand{\NCSA}{National Center for Supercomputing Applications, Urbana, IL, 61801, USA}
\newcommand{\NSF}{National Science Foundation Graduate Research Fellow}
\newcommand{\CCA}{Flatiron Institute Center for Computational Astrophysics, New York City, NY 10010, USA}
\newcommand{\CMU}{McWilliams Center for Cosmology, Department of Physics, Carnegie Mellon University, Pittsburgh, PA, USA}

\begin{document}

\title{\vspace{-3em} First Impressions: Early-Time Classification of Supernovae using Host Galaxy Information and Shallow Learning}
\author[0000-0003-4906-8447]{Alexander~Gagliano}\blfootnote{Corresponding author: Alex~Gagliano\\ \href{mailto:gaglian2@illinois.edu}{gaglian2@illinois.edu}}
\affiliation{\Illinois}
\affiliation{\NCSA}
\affiliation{\CCA}
\affiliation{\CAPS}
\affiliation{\NSF}

\author[0000-0002-3011-4784]{Gabriella~Contardo}
\affiliation{\CCA}

\author[0000-0002-9328-5652]{Daniel~Foreman~Mackey}
\affiliation{\CCA}

\author[0000-0002-8676-1622]{Alex~I.~Malz}
\affiliation{\CMU}

\author[0000-0002-6298-1663]{Patrick~D.~Aleo}
\affiliation{\Illinois}
\affiliation{\NCSA}
\affiliation{\CAPS}

\begin{abstract}
Substantial effort has been devoted to the characterization of transient phenomena from photometric information. Automated approaches to this problem have taken advantage of complete phase-coverage of an event, limiting their use for triggering rapid follow-up of ongoing phenomena. In this work, we introduce a neural network with a single recurrent layer designed explicitly for early photometric classification of supernovae. Our algorithm leverages transfer learning to account for model misspecification, host galaxy photometry to solve the data scarcity problem soon after discovery, and a custom weighted loss to prioritize accurate early classification. We first train our algorithm using state-of-the-art transient and host galaxy simulations, then adapt its weights and validate it on the spectroscopically-confirmed SNe~Ia, SNe~II, and SNe~Ib/c from the Zwicky Transient Facility Bright Transient Survey. \textcolor{black}{On observed data,} our method achieves an overall accuracy of \textcolor{black}{$82 \pm 2$\%} within 3 days of an event's discovery, and an accuracy of \textcolor{black}{$87\pm5$\%} within \textcolor{black}{30} days of discovery. At both early and late phases, our method achieves comparable or superior results to the leading classification algorithms with a simpler network architecture. These results help pave the way for rapid photometric and spectroscopic follow-up of scientifically-valuable transients discovered in massive synoptic surveys.%, to ensure high-cadence photometric and spectroscopic coverage. %, purity of SOMETHING%, and a completeness of SOMETHING% within the first day of trigger for a sample of spectroscopically-classified events. We discuss the scalability of our method and the value of host-galaxy information in classifying more exotic transients, including Tidal Disruption Events.
 
\end{abstract}

\section{Introduction}
\label{intro} 
Decades of spectroscopic insights have motivated the construction of a hierarchical classification taxonomy for the death knells of stars as supernovae \citep[SNe;][]{1997Filippenko_Classification, 2017Galyam_Classification}. At its core are the four most commonly-observed classes: Type-Ia (SNe~Ia); Type-II (SNe~II); and Type-Ib and Type-Ic (collectively SNe~Ib/c). SNe~Ia, signposts for the thermonuclear detonations of carbon-oxygen white dwarfs, are defined by the early presence of strong Si II features. SNe~II, the explosions of young, massive stars (with a Zero-Age-Main-Sequence mass of $M_{\rm ZAMS} > 8\; M_{\odot}$) following core collapse, are events whose spectra exhibit strong hydrogen lines with P-Cygni profiles. The less-common SNe~Ib/c \citep[comprising, together with the transitional SNe IIb, $\sim$30\% of all core-collapse events;][]{2017Shivvers_SNRates} exhibit spectra devoid of both hydrogen and Si II features; these events result from the core-collapse of young stars that have undergone envelope stripping.

Despite together comprising the vast majority of observed terminal explosions, SNe~Ia, II, and Ib/c remain poorly understood in both their detonation mechanisms and the physics driving their observed diversity \citep{2021Laplace}. SNe~Ia are likely triggered by either the runaway burning of an accreting white dwarf pushed beyond the Chandrasekhar limit by a close-in massive companion (with significant uncertainty surrounding the nature of the companion; see \citealt{1999Hachisu_SingleDegenerate}), or the merger of two white dwarfs \citep{2012Pakmore_SNIa_DoubleDegenerate}. Significant heterogeneity exists among SNe~II in both their photometric evolution \citep{2012Arcavi_CCCP} and spectroscopic characterization \citep{2013Taddia_CSP} that has yet to be connected definitively to the nature or pre-explosion behavior of their progenitors. Stripped-envelope SN (SN~Ib/c/IIb) progenitors likely arise preferentially in binaries with some degree of mass transfer \citep{1992Podsiadlowski_BinaryInteraction}, although some are believed to lose their envelopes in isolation through eruptive mass-loss and metallicity-driven stellar winds \citep{2018Kuncarayakti_CCProgenitors}. Further, observations have been unable to clarify whether SNe~Ib, whose spectra contain clear helium signatures, and SNe~Ic, whose spectra lack them, arise from distinct stripping channels or instead reflect extremes along a continuum in stripping degree. % The presence of these events in both early and late-type galaxies (CITE) and the discovery of multiple sub-types in recent years (e.g., SNe~Iax; Foley) suggests that multiple channels are likely, although the remarkable homogeneity of the Branch-normal class (which has been the cornerstone of cosmological advances; CITE) suggests one as the dominant mechanism (CITE).

Although their underlying physics remain unclear, we are now discovering these transients in abundance. Time-domain surveys, in an attempt to balance sky coverage and depth, can be broadly distinguished between those that are low-redshift and wide-field, such as the Nearby SN Factory \citep[SNFactory;][]{2002Algering_SNFactory}, the Texas SN Search \citep{2006Quimby_TSS}, the Pan-STARRS Survey for Transients \citep[PSST;][]{2015Huber_PSST}, the All-Sky Automated Survey for SNe \citep[ASAS-SN;][]{2017Kochanek_ASASSN}, the Asteroid Terrestrial-impact Last Alert System \citep[ATLAS;][]{2018Tonry_ATLAS}, the Zwicky Transient Facility (ZTF; \citealt{2019Bellm_ZTF}), and the Young SN Experiment \citep[YSE;][]{2021Jones_YSE}; and those that are targeted, high-redshift, and narrow-field, including the Sloan Digital Sky Survey-II SN Survey \citep{2008Sako_SDSSII}, SN Cosmology project \citep{1999Perlmutter_SNCosmology}, Deep Lens Survey \citep{2004Becker_LensSurvey}, and the Pan-STARRS Medium Deep survey \citep[MDS;][]{2019Villar_MDSClassification}; among others. These complementary searches have contributed to an exponentially-increasing ($\mathcal{O}(10^4\;\rm{yr}^{-1})$) SN discovery rate. The Vera Rubin Observatory \citep[Vera Rubin Obs;][]{2019Ivezic}, with its synoptic coverage (completely covering the Southern sky each $3-4$ days in its Wide-Fast-Deep mode) and its unprecedented survey depths ($r\sim24.5$ in a single visit), is slated to annually discovery $10^6$ transients\footnote{\href{https://www.lsst.org/sites/default/files/docs/sciencebook/SB_11.pdf}{https://www.lsst.org/sites/default/files/docs/sciencebook/SB_11.pdf}}, single-handedly surpassing the current exponential scaling of SN discovery.

SNe occur at the interface between a dying progenitor star and its post-explosion remnant. While we are unable to witness the exact moment of this metamorphosis, at best we can deepen our understanding by tightening our observations in both temporal directions: in other words, by observing and connecting the final moments of the progenitor star's life and the earliest moments of its destruction. Following this approach, recent high-cadence imaging has revealed multiple serendipitous detections of pre-cursor emission in the weeks to months preceding the explosions of core-collapse SNe (e.g., 2009ip, 2010mc, 2020tlf;  \citealt{2013Mauerhan_2009ip},  \citealt{2013Ofek_Precursor}, and  \citealt{2022Jacobson_2020tlf}, respectively). The depth of LSST imaging offers a promising avenue for detecting progenitor emission at \textit{some} pre-explosion phase, but the synoptic coverage of this and other upcoming surveys comes at the expense of sparse phase coverage of both any pre-explosion activity and of the terminal explosion. Our capacity to classify discovered events is further thwarted by our lack of complementary spectroscopic resources. Data from LSST and other upcoming surveys must be rapidly augmented by additional follow-up from targeted instruments to fully characterize the photometric and spectroscopic evolution of an event and link an explosion to its pre-cursor activity, particularly at early phases when the physics of an explosion is most directly encoded onto its light curve.

\textcolor{black}{To consolidate photometrically pure samples, and} to facilitate the follow-up of scientifically valuable events, light curve classifiers have appeared on the scene en masse. Some of these require matching observed light curves to archival templates \citep[e.g.,][]{2011Sako_PSNID}; others classify on features extracted from light curves (either properties of the rise, peak, and decline as estimated with parametric fits, or reduced-dimensionality representations of the full light curve; see \citealt{2016Lochner_PhotometricClassification} and \citealt{2019Villar_MDSClassification} for reviews of multiple popular techniques). Several of these algorithms are founded upon principles directly relevant to solving the early-time classification problem. These principles include: \textit{speed}, with 15M light curve classifications per second processed with the neural network \texttt{SuperNNova} \citep{moller2020supernnova}; \textit{flexibility}, with the auto-encoder and recurrent neural network \texttt{SuperRAENN} \citep{2020Villar_superRAENN} able to develop realistic light curve representations for classification without complete phase coverage; and \textit{adaptive prediction}, first shown as a proof-of-concept by \cite{muthukrishna2019rapid} with their recurrent network \texttt{RAPID} \textcolor{black}{and more recently extended to \textcolor{black}{more sophisticated} network architectures in \cite{2022Scone_PhotometricClassification} and \cite{2022Pimentel_DeepSNClassification}}. 

Despite these early successes, real-time transient classification remains in its infancy. Success in practice has been limited by two factors. First, to satiate the data-starved machine learning classifiers currently in use, it has become standard practice to generate and train on a large sample of synthetic light curves. Due to our limited understanding of the underlying physical systems involved, these transient models are driven by observed phenomenology, and so are necessarily simplified. This limits the performance of these classifiers on observed samples, particularly those containing out-of-distribution behavior. Second, the question of early-time classification has to date been explored \textit{post-hoc} through modifications to networks originally developed for full-phase classification. A classifier has yet to be designed explicitly to classify early and to bridge the divide spanning simulation and observation.

The insights to be gleaned from high-cadence early-time coverage of SNe, even common ones, are vast. Early-time brightness variations can reveal interactions with a companion star in a binary, a surrounding progenitor envelope, or circumstellar material shed by the progenitor pre-explosion \citep[e.g.,][]{2010Kasen_Companion, 2008Gezari_SBO, 2019Dimitriadis_2018oh, 2019Shappee_SeeingDouble, 2022Gagliano_2020oi}, and comparisons with analytic models of $^{56}$Ni decay can reveal tantalizing evidence for additional engines powering an explosion (as is the case for SLSNe-I and, recently, SNe Ic; see \citealt{2022Hosseinzadeh_SLSNBumps} and \citealt{2021Afsariardchi_IcPowerSources}, respectively). The evolution of an explosion’s color in the first few days, as discrete tracers of its underlying spectral energy distribution (SED), can even unveil an asymmetric distribution of nucleosynthetic products, such as an overabundance of heavy elements on the surface of the progenitor star \citep{2022Ni_Infant}. 

Real-time classification will also help enable the discovery of a greater number of rare and rapidly-evolving transients that have eluded discovery within previous surveys. Serendipitous detections among massive survey streams have already revealed multiple transients in this region of parameter space, including Rapidly-Evolving Transients (RETs; \citealt{2018Pursiainen_RETs}); Fast Blue Optical Transients (FBOTs; \citealt{2019Perley_COW}); and Fast-Evolving Luminous Transients (FELTs; \citealt{2018Rest_FELTs}). Because the emission timescales of most SNe are determined by the radioactive half-life of $^{56}$Ni, events evolving more rapidly than this hint at distinct explosion physics and central engines (such as a relativistic jet that interacts with a cloud of dense circumstellar material, as with FBOTs; \citealt{2022Gottlieb_FBOTJets}). Increasing our sample sizes of these events is essential for constructing a complete picture of stellar death and the nature of the compact remnant that these doomed systems leave behind.

In this work, we introduce a recurrent neural network designed end-to-end for \textit{early} classification of explosive transients. In contrast to the rise of deep-learning methods that introduce complex networks at the expense of model interpretability, our model consists of a single recurrent layer and leverages physical information about the explosion site to improve early performance. Because of the simplified architecture and the emphasis on classification with incomplete light curve information, we deem this approach `shallow learning'. Our framework extends the initial real-time classification system developed by \cite{muthukrishna2019rapid} in four key respects: 
\begin{enumerate}
    \item We implement a temporally-weighted categorical cross-entropy loss function such that early classification is prioritized over full-phase classification. 
    \item We adopt a flexible Gaussian Process model parameterized by the correlation between filter passbands and observations in time to realistically interpolate observed transient photometry.
    \item We utilize a two-component training strategy where the network is first trained on a large sample of simulated light curves, and then re-trained on a curated subset of real observations. 
    \item We include photometry from the galaxy where a transient occurred (its `host galaxy') to reduce the reliance on sparse transient photometry at early phases.
\end{enumerate}
Host galaxy information is increasingly being considered for its value in early SN classification. The initial work of \cite{foley2013classifying} verified using the Lick Observatory SN Search sample \citep{2001Filippenko_LOSS} that host galaxy morphology and color could be used to construct photometric SN~Ia samples as pure as those constructed using the then-leading light curve methods. \cite{2020Baldeschi_PS1} extended this work with a random forest classifier to distinguish between low and highly star-forming host galaxies, finding that this metric could serve as an indicator of transient type in the two-class scenario (SNe~II versus SNe~Ia). A postage-stamp classifier constructed by \cite{2021Carrasco-Davis_PostageStamp} uses a convolutional neural network to distinguish between SNe, asteroids, variable stars, and bogus alerts with a single detection image (in the case of SNe, this necessarily includes host galaxy information). \cite{2021Gagliano_GHOST} found that a random forest classifier trained directly on the photometric features of transient host galaxies achieved $\sim70\%$ accuracy distinguishing between SNe~II and SNe~Ia. FLEET, a random-forest classifier used to recover SLSNe-I from survey streams, combines a set of full-phase parametric light curve features with host galaxy properties and reports 20 SLSN-I discovered per year with an overall detection purity of 85\% \citep{2020Gomez_FLEET}. More recent work has extended these insights to additional sub-classes of SNe \citep{2022Kisley_OnlyHostPhotometry}. None of these methods leverage both host-galaxy and light curve information for \textit{real-time}, \textit{adaptive} photometric classification.

We leverage the photometric information from two surveys in this work, mimicking the realistic scenario in which heterogeneous data (potentially with observations collected in distinct filter systems) \textcolor{black}{are} consolidated from multiple sources. We consider SN photometry in ZTF-$gr$ and $grizy$ host-galaxy photometry \textcolor{black}{from the Pan-STARRS 3-$\pi$ survey \citep{chambers2016pan}}, both in our simulated and observed samples.

\begin{figure*}
    \centering
\includegraphics[width=\linewidth]{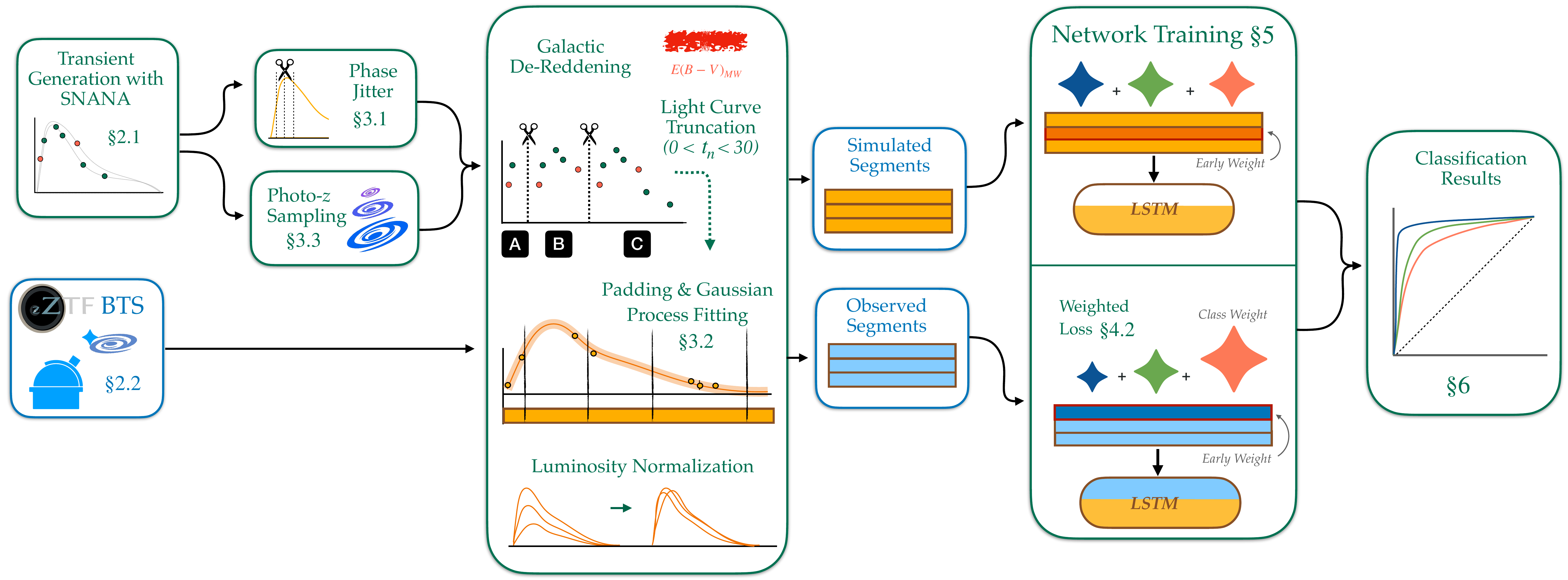}
    \caption{Overview of our analysis pipeline, with relevant sections in this paper labeled. Data are outlined in blue, while analysis steps are given in green.}
    \label{fig:overview}
\end{figure*}

We provide an overview of our analysis, which mimics the structure of this paper, in Fig.~\ref{fig:overview}. In \textsection\ref{sec:data}, we introduce the simulated and observational datasets used in this work. We outline our prescription for pre-processing each of the transient light curves in \textsection\ref{sec:preprocess}, and introduce our recurrent neural network architecture in \textsection\ref{sec:architecture}. We then describe our training of the model in \textsection\ref{sec:training}. We provide results from our recurrent classification tests and place them in the context of \textcolor{black}{light curve only classification, and of} other recently-released photometric classifiers\textcolor{black}{,} in \textsection\ref{sec:results}. We conclude with a discussion of science applications and future work in \textsection\ref{sec:conclusions}. 

\section{Data}\label{sec:data}

\subsection{Simulated Transients and their Host Galaxies}\label{subsec:sim_lcs}

\subsubsection{Modeling Transient Photometry with SNANA}\label{subsubsec:simulated_transients}
\textcolor{black}{We first outline the pipeline used to generate the primary training set for our classification network. Our strategy leverages the popular simulation code} \texttt{SNANA}\textcolor{black}{\footnote{https://github.com/RickKessler/SNANA/tree/master}} \citep{Kessler2009_SNANA}, which constructs a forward model for a transient event starting from a rest-frame SED and ending with synthetic observations matching the cadence and depth of a survey of interest. The simulation considers atmospheric distortions, instrument efficiency, and photometric calibration in estimating the uncertainty of each simulated observation. A common set of \textcolor{black}{SN SED} models used to train the current generation of photometric classifiers are those constructed for the Photometric LSST Astronomical Time-Series Classification Challenge \citep[PLAsTiCC;][]{2019Kessler_PLAstiCC}. Since the challenge was first run, the models associated with SNe~Ia, SNe~II, and SNe~Ib/c have all been updated, and will soon be released in an updated iteration of the challenge. \textcolor{black}{We adopt these updated SED templates for our simulations.}

\textcolor{black}{Using \texttt{SNANA}, we} simulate the properties of four classes of SNe: SNe~Ia, SNe~II, SNe~Ib, and SNe~Ic. We assume a flat $\Lambda$CDM cosmology with $H_{0}$ = 70 km s$^{-1}$ Mpc$^{-1}$, $\Omega_{M}$ = 0.3, $\Omega_{\Lambda}$ = 0.7, and $w = -1$. To parameterize the rise, stretch, and color of SNe~Ia as a function of their rest-frame SEDs, we use the recently-released \texttt{SALT3} model \citep{Kenworthy2021_SALT3}, an update of the commonly-used \texttt{SALT2} model \citep{SALT22007}. For SNe~II,  SNe~Ib, and SNe~Ic, we use the spectrophotometric templates provided by \cite{Vincenzi2019_Templates}. These templates have been constructed from observed events of multiple sub-types; we consolidate the templates provided for SNe~IIP/IIL, SNe~IIn, and SNe~IIb as ``SNe~II'' to reflect the observational diversity of this class. 

We generate synthetic light curves in ZTF-$g$ and ZTF-$r$ for each transient class over the redshift range $0<z<0.2$ to match the photometric properties of the data publicly reported in the alert stream of the Zwicky Transient Facility \cite[ZTF;][]{2019Bellm_ZTF}\footnote{\url{http://ztf.caltech.edu}}. 
 Because we aim to apply our trained network to the ZTF Bright Transient Survey (ZTF BTS), which is magnitude-limited to ZTF-$r<18.5$ (additional detail on the catalog is provided in \textsection\ref{subsec:ZTFBTS}), we cut from our simulations any transients without at least a single observation brighter than this limit. We also remove observations with a signal-to-noise ratio (S/N) of less than 4 (determined empirically to match the S/N distribution of observations reported in the ZTF BTS). The sky locations of transients in the simulated sample, which convolves a uniform distribution where they occur with the ZTF footprint where they are detected, are then used to calculate the expected line-of-sight Galactic extinction assuming $R_V = 3.1$, an intrinsic scatter of $\sigma_{E(B-V)} = 0.16$, and the dust law from \cite{1989Cardelli_Dust}.
 
\textcolor{black}{Because the simulation represents a forward model imposing observational cuts on an underlying `true' transient sample, the final number of transients `observed' is unknown \textit{a priori}. We simulate 50,000 events of each class so that at least 20,000 remain after these cuts.} Once we have a dataset of `observed' ZTF SNe, we combine SNe~Ib and SNe~Ic into a single ``SN~Ib/c'' class (due to the uncertainty surrounding their classification as distinct sub-types) and under-sample the dominant classes with the Python package \texttt{imbalanced-learn} \citep{JMLR:v18:16-365} so that the full dataset has the same number of events per class. This results in a sample containing 20,000 ZTF light curves per transient class (SN~Ia, SN~II, and SN~Ib/c), or 60,000 in total.

We have simulated our transients using 10 Intel Xeon ``Haswell” processor nodes on the Cray XC40 ``Cori” system at the National Energy Research Scientific Computing (NERSC). The full simulation took a total wall time of 5.2 hours. 

\subsubsection{Simulated Host-Galaxy Properties with Normalizing Flows}\label{subsubsec:simulated_galaxies}
\textcolor{black}{Our goal is to leverage host-galaxy properties to predict a transient's class when explosion photometry is limited and spectroscopy is absent. Our neural network needs to be trained on data comprising only what is available at test-time, and should be informed by extant SN samples.}

\textcolor{black}{Significant effort has been devoted to expanding the \texttt{SNANA} simulation code to simultaneously model samples of observed transients and their host galaxies \citep{2019Brout_FirstCosmology,Vincenzi2021,2022Lokken_SCOTCH}; however, this approach has several limitations for our use case. First, the generation of realistic photometric redshifts demands that both the survey-specific photometry and colors of synthetic galaxies be well-calibrated against observations \citep[a non-trivial problem; ][]{CosmoDC22019}. Second, we wish to generate a large, balanced training sample spanning the same redshift range as the test set. This would likely require scaling up the number of low-redshift galaxies in existing synthetic or observed catalogs, which can introduce artificial structure if properties are repeated. Finally, as ZTF photometry does not exist for galaxies, our framework needs to match SNe observed in one survey to galaxies observed in another.}

%These \cite{2022Lokken_SCOTCH} considered the redshift range spanned by all transients discovered by the Vera C. Rubin Observatory ($0<z<3$), and photometric redshifts have recently been simulated from deep co-adds from the Vera C. Rubin Observatory. Though these data products will be invaluable for upcoming analyses, our strategy is to optimize our network for \textit{the current generation} of low-$z$ SN discovery. 

\textcolor{black}{We bypass these limitations by proposing a simple data-driven model trained to reproduce the class-specific photometric properties (PS1-$grizy$ and photometric redshift) of observed SN host galaxies. By conditioning these properties on the true redshifts of our generated transients, we can ensure that the host galaxy properties are consistent with the redshift range simulated without repeating values or demanding that they be observed in the same survey. To this end, we train separate conditional normalizing flows for the host-galaxy properties of each transient class in our sample (SN~Ia, SN~Ib/c, and SN~II).}

Normalizing flows \citep{2015Jimenez_NormalizingFlow} are generative models constructed to approximate and draw from a joint probability density function (PDF) associated with a complex multi-dimensional dataset. They are parameterized by an invertible bijective function (or bijector) that maps a simple distribution (e.g., a multi-dimensional Uniform distribution) to the observed data distribution. New samples can then be drawn from the data distribution by sampling the simple distribution and applying the learned bijective function. A flow can also be constructed to predict multiple parameters of the dataset conditioned on a separate parameter, as is done here. \textcolor{black}{When trained on observed events, this approach allows us to generate realistic host-galaxy properties without an underlying galaxy model or an encoding of the observing systematics limiting their identification.}  We use the \texttt{PZFlow}\footnote{\url{https://github.com/jfcrenshaw/pzflow}} package \citep{2022PZFlow_JFC} to build our conditional normalizing flow.

\textcolor{black}{We first consolidate PS1 best-fit Kron magnitudes in Pan-STARRS $grizy$-passbands for the galaxies within the GHOST \citep{2021Gagliano_GHOST} catalog. After excluding events from the ZTF BTS sample (described in the following section), those without spectroscopic redshift information, and those without complete PS1-$grizy$ photometry, we are left with $\sim8,000$ events.  The PS1 photometry for each galaxy is corrected for Galactic extinction using the total reddening values at each location derived from Planck observations of the cosmic microwave background \citep{2014Planck_Cosmo}. We \textcolor{black}{further consolidate} photometric redshifts for each host galaxy from the neural-network-produced `PS1-STRM' catalog \citep{2021Beck_pzLSTM}, and separate `SNe Ia', 'SNe II', and `SNe Ib/c' hosts.}

% We accomplish these goals by constructing a normalizing flow conditioned on PS1-$grizy$ photometry.

\textcolor{black}{We then train a conditional flow to generate extinction-corrected PS1-$grizy$ photometry and photometric redshifts for each transient type given the transient's spectroscopic redshift. Our bijective function is a Rational-Quadratic Neural Spline Coupling \citep{2019Durkan_NSC}, and our target distribution is a multi-dimensional Uniform distribution. We train for 100 epochs and confirm convergence of the log-probability. We then condition our learned flow on the spectroscopic redshift of each \texttt{SNANA}-simulated transient to generate a posterior PDF for its host galaxy properties: $p(g,r,i,z,y,z_{\rm{phot}}|z_{\rm{spec}})$. We draw from this joint PDF to simultaneously generate point estimates for each of these properties. We note that uncertainties for these properties can be estimated by repeatedly drawing from this conditional flow, or the full posterior can be used instead; for this work we limit ourselves to the single estimate of each property.}

\textcolor{black}{We emphasize that we have not implemented any host-galaxy association technique, as no positional information exists: our simulated hosts are fully described by the properties we have generated with the normalizing flow. In practice, SNe observed within $z<0.6$ may be misassociated at the 3-5\% level using traditional techniques \citep{2021Gagliano_GHOST}; reproducing this contamination rate using simulations requires an all-sky galaxy catalog with realistic number densities, far beyond the scope of this work. Because this contamination induces additional scatter in the observed host-galaxy correlations for each SN class, our normalizing flow trained on observations naturally encodes the effects of these misassociations in our simulated galaxy photometry (though misassociations are less likely within our considered redshift range of $z<0.2$).}

%\textcolor{black}{The SCOTCH \citep{2022Lokken_SCOTCH} framework within \texttt{SNANA} to realistically associate simulated transients with host galaxies. SCOTCH consists of a series of host galaxy libraries informed by observed samples (e.g., \citealt{2021Gagliano_GHOST}), best-fit scaling relations for the host galaxies of different transient types (e.g., \citealt{Vincenzi2021}), and the synthetic sky catalog CosmoDC2 \citep{CosmoDC22019}.}

\textcolor{black}{Next, we} divide our simulated sample into training and test sets comprising 75\% and 25\% of the original dataset, respectively. We maintain the even class balance in the training set; however, we re-balance the test set to have the approximate class proportions as exist in the ZTF BTS after quality cuts: 80\% SNe~Ia, 15\% SNe~II, and 5\% SNe~Ib/c (where we assume equal proportions of SNe~Ib and SNe~Ic in the combined SN~Ib/c class). This allows us to track the performance of the network on a BTS-like sample of events as it trains. We achieve these proportions by undersampling our SN~II and SN~Ib/c light curves using \texttt{imbalanced-learn}. Our final simulated training set contains 10,000 events of each class, while our simulated test set contains 8,000 SNe~Ia, 1,500 SNe~II, and 500 SNe~Ib/c. 

%When simulating these transients, we adopt the same transient host-galaxy correlations as were used in SCOTCH, which include photometric and derived correlations that have been validated against observed samples \citep{2022Lokken_SCOTCH}. In the simulation code, where host galaxies are associated with transients based on their properties, their observed photometry in LSST $ugrizY$ passbands is retrieved from the CosmoDC2 catalogs. For adaptation to PS1 $griz$ bands, we use the provided LSST $griz$ photometry in our training. We construct an adaptive training stage, which we discuss in greater detail in \textsection \ref{retraining}, for the network to shift from using the simulated LSST photometric system to using the observed YSE photometric system.

\subsection{Observed Transients from The ZTF Bright Transient Survey (ZTF BTS)}\label{subsec:ZTFBTS}
To evaluate the performance of our network on real observations, we use the set of \textcolor{black}{4,020 high-quality}\footnote{\textcolor{black}{See \url{https://sites.astro.caltech.edu/ztf/bts/explorer_info.html} for a description of quality and purity cuts.}} SNe from the ZTF BTS\footnote{\href{https://sites.astro.caltech.edu/ztf/bts/bts.php}{https://sites.astro.caltech.edu/ztf/bts/bts.php}} \citep{2020Fremley_ZTFBTSI, 2020Perley_ZTFBTSII}, the largest spectroscopic sample of SNe constructed to date. Data from the ZTF public stream are collected using the 47~deg$^{2}$ field-of-view camera mounted atop the Palomar 48-inch Schmidt telescope, which observes in three passbands: ZTF-$g$, ZTF-$r$, and ZTF-$i$. The data collected, which now include forced photometry at the locations of identified events \citep{2019Masci_ZTFForced}, are then processed at the Infrared Processing and Analysis Center (IPAC) to produce public alerts $\sim$4 minutes from the time of observation. Now in the second phase of its public survey, which comprises half of its total time on-sky, ZTF scans the entire northern sky with a cadence of $\sim$2 nights in ZTF-$g$ and ZTF-$r$.

The ZTF BTS is magnitude-limited to ZTF-$r<18.5$ and nearly spectroscopically-complete, containing spectroscopic classifications for 93\% of discovered SNe. The catalog (at the time of download on March 16th, 2023) consists of \textcolor{black}{3,275 SNe~Ia, 563 SNe~II, and 182 SNe~Ib/c}, whose photometry was released to alert brokers in near real-time and whose spectra were made public on the Transient Name Server\footnote{\href{https://www.wis-tns.org}{https://www.wis-tns.org}} within a day of observation. The catalog also contains multiple events belonging to rarer classes (including SLSNe-I); we do not consider these additional classes in this study, but note that significant host-galaxy correlations have been identified among these other classes \citep{2022Lokken_SCOTCH} and our model can easily be extended to include the\textcolor{black}{m}. %'SN Ia': 2788, 'SN II': 493, 'SN Ib/c': 163}

We identify the most likely host galaxies of these transients in Pan-STARRS (PS1) using the modified directional light-radius \citep{gupta2016host} method implemented in the GHOST \citep{2021Gagliano_GHOST} software\footnote{\href{https://pypi.org/project/astro-ghost/}{https://pypi.org/project/astro-ghost/}}. The host galaxies of 866 SNe could not be found using this method and were dropped. Through visual inspection, another 76 were found to have erroneous or ambiguous associations and \textcolor{black}{were manually re-associated}.%removed from the sample. 

For the identified hosts, \textcolor{black}{PS1 best-fit Kron magnitudes in $grizy$-passbands were retrieved. 301 systems had unreported host galaxy photometry in at least one PS1 band and were dropped at this stage, leaving 2,853 events.} We then downloaded the photometry of these transients using the ANTARES alert broker local client\footnote{\href{https://nsf-noirlab.gitlab.io/csdc/antares/client/installation.html}{https://nsf-noirlab.gitlab.io/csdc/antares/client/installation.html}} \citep{2021Matheson_ANTARES}. \textcolor{black}{Photometry for 2 transients could not be found and these events were removed.} 

%These magnitudes are determined by integrating the best-fit brightness profile of each galaxy, where the best-fit model is determined in SDSS-$r$ between a de Vaucouleurs and an exponential model\footnote{\href{https://www.sdss.org/dr12/algorithms/magnitudes/\#mag\_model}{https://www.sdss.org/dr12/algorithms/magnitudes/\#mag_model}}. Because the PS1 survey spans more of the sky than SDSS, we use linear regression to determine a mapping between PS1-$griz$ Kron magnitude (the brightness contained within an aperture twice the size of the first image moment radius) and SDSS-$griz$ model magnitude for the galaxies spanning both surveys. We use this function to compute the expected SDSS-$griz$ photometry for each host galaxy observed only in PS1. 

\textcolor{black}{Extensive prior work has been done to construct and validate photometric redshift estimators for specific surveys and galaxy datasets. Due to the lack of generalizability of these methods, many surveys report neural-network generated photometric redshifts for each galaxy but not the associated code used to generate them. To bypass this issue, we have constructed a simple emulator for the photometric redshifts reported in \cite{2021Beck_pzLSTM} for transient host galaxies. This allows us to recover redshifts with comparable statistical properties to those predicted by fine-tuned models.}

Similar to the approach for the simulated sample, we train a conditional flow \textcolor{black}{to generate data with comparable statistical properties to our observed sample. Whereas our previous flow was used to generate both host-galaxy photometry \textit{and} photometric redshifts for the simulated sample, this model leverages observed host-galaxy photometry to predict \textit{only} the \cite{2021Beck_pzLSTM}-reported photometric redshifts}. We condition our flow on PS1-$r$ and the colors of galaxies within the GHOST \citep{2021Gagliano_GHOST} sample, which correlate more strongly with redshift than galaxy brightness alone. \textcolor{black}{As a result, our normalizing flow predicts the conditional distribution} $p(z_{\rm{phot}}|r,g-r,r-i,i-z,z-y)$. As before, we train the flow for 100 epochs and verify convergence of the log-probability.

\textcolor{black}{After training, we apply our normalizing flow to our ZTF BTS host galaxy photometry to recover photometric redshift point estimates. We use these values} to normalize our light curves as described in the following section. We similarly use the PS1 zero-points and photometric redshifts to convert the host galaxy apparent magnitudes to host galaxy luminosities, which we normalize to near unity. 

\begin{figure}
    \centering
    \includegraphics[width=\linewidth]{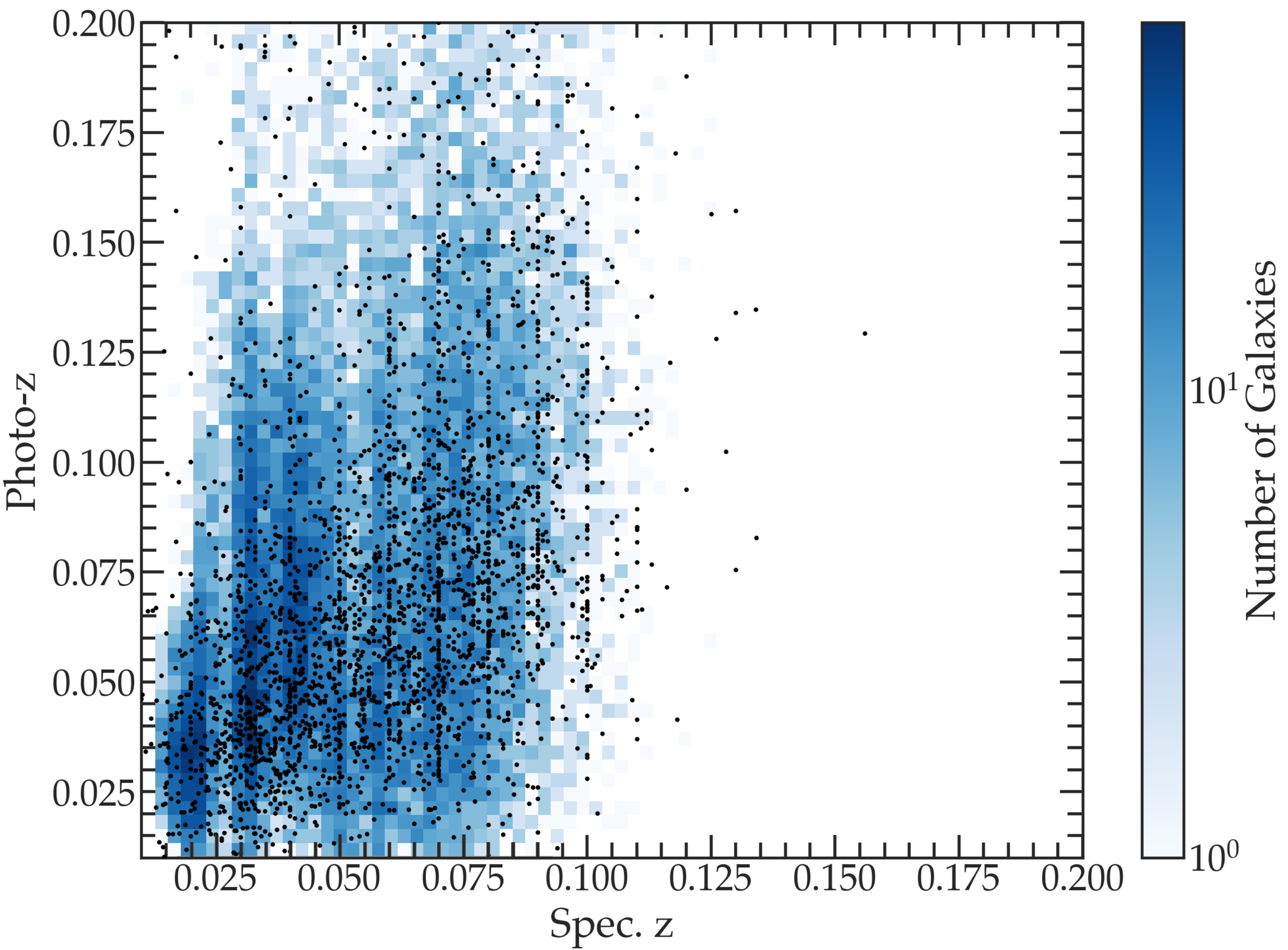}
    \caption{Photometric versus spectroscopic redshifts for the galaxies in our simulated and observed samples. ZTF BTS host galaxies are shown as black points, and synthetic host galaxies are shown as the two-dimensional histogram in \textcolor{black}{blue. The discreteness in observed spectroscopic redshifts is an artifact of the template-matching algorithm through which they were determined.}}
    \label{fig:photo_zs}
\end{figure}

We plot the distribution of photometric redshifts \textcolor{black}{for both the simulated and observed samples} as a function of \textcolor{black}{their spectroscopic redshifts} in Fig.~\ref{fig:photo_zs}. Qualitatively, the scatter in photometric redshifts between the two samples is similar. \textcolor{black}{Quantitatively, we can measure the number of photometric redshift outliers as \citep{2010Hildebrandt_PHotoz}
\begin{equation}
\frac{|z_{\rm{phot}}-z_{\rm{spec}}|}{1+z_{\rm{spec}}} > 0.15
\end{equation}
By this metric, we measure an outlier fraction of 4.5\% for the observed sample and 1.5\% for the simulated sample (compared to 2.5\% for the \cite{2021Beck_pzLSTM}-reported photometric redshifts for the GHOST training sample).}

%We caution that these models have been constructed using mutually non-representative photometry from different surveys (LSST's 10-year co-added image depth is $r\sim$26.5 compared to PS1's $r\sim$23.2\footnote{\href{https://outerspace.stsci.edu/display/PANSTARRS/}{https://outerspace.stsci.edu/display/PANSTARRS/}}), and a more thorough treatment would aim to adapt the normalizing flow to more closely emulate the statistical properties of the redshifts predicted by \texttt{Easy PhotoZ}. Nevertheless, 
\textcolor{black}{We note that the use of $u$-band photometry from the Sloan Digital Sky Survey (SDSS; \citealt{2000York_SDSS}) would likely decrease this scatter, but the increased depth of PS1 results in higher-S/N host-galaxy photometry for use in our network. Further, redshift is a key feature used to estimate the
absolute brightness of an explosion, and therefore a key discriminator between SNe Ia and core-collapse events. The reliability of photometric redshifts in the era of the Vera C. Rubin Observatory is an area of active research, and we avoid fine-tuning these estimates further so as not to report overly-optimistic performance. Our approach represents a significant advancement over previous light curve simulations that estimate photo-zs by interpolating look-up tables of host galaxy colors \citep[see e.g.,][]{2018Graham_LSSTPhotozs}.} %\textbf{DO WE NEED TO INCLUDE THIS ANYMORE??}

Our observed sample before pre-processing contains 2,851 SNe. An additional 44 were dropped during the light curve truncation stage described in \textsection \ref{sec:preprocess}, leaving us with a final observed sample size of 2,807 SNe. While this is a fraction of the initial number in the ZTF BTS ($\sim$70\%), we note that only a small fraction of the discovered SNe in upcoming surveys will be classified and selected for follow-up observations. By constructing a curated observational dataset, we optimize our classifier's ability to discover events for which we will be able to extract meaningful scientific insights. We compare the distributions of peak apparent magnitude, redshift, and host galaxy photometry after each of these \textcolor{black}{preprocessing} steps to ensure that we do not encode an additional observational bias into our observed sample with these cuts. We summarize our sample sizes after each of these cuts in Table~\ref{tbl:cuts}.

\begin{table}[]
    \centering
    \begin{tabular}{c|c|c}
         Processing Cut & SNe Remaining & Fraction Cut  \\ \hline \hline
         \textcolor{black}{Initial Sample} & \textcolor{black}{4,020} & --\\
      %  Missing Photometry & 3,436 & 0.2\% \\
        \textcolor{black}{Missing/Incorrect Host} & \textcolor{black}{3,154} & \textcolor{black}{21.5}\% \\
        %Incorrect Host & 3,113 & X.XX\% \\
        \textcolor{black}{Unreported Host Phot.} & \textcolor{black}{2,853} & \textcolor{black}{9.6}\% \\
          \textcolor{black}{SN Light Curve Retrieval} & \textcolor{black}{2,851} & \textcolor{black}{$<0.1$}\% \\
        \textcolor{black}{Light Curve Truncation} & \textcolor{black}{2,807} & \textcolor{black}{1.5}\% \\ \hline 
        \textcolor{black}{Final SN Sample} & \textcolor{black}{2,807} & \textcolor{black}{30.2}\% \\
    \end{tabular}
    \caption{Number of SNe from the ZTF BTS remaining, and fraction removed, after each step in the processing pipeline.}
    \label{tbl:cuts}
\end{table}

We divide our final spectroscopic sample into a 75\% training set and a 25\% test set without changing the relative proportions of classes. As a result, \textcolor{black}{our spectroscopic training set consists of 1,704 SNe~Ia, 280 SNe~II, and 121 SNe~Ib/c; our spectroscopic test set consists of 569 SNe~Ia, 95 SNe~II, and 38 SNe~Ib/c.}

\section{Data Pre-Processing}\label{sec:preprocess}

\begin{figure*}
    \centering
    \includegraphics[width=\linewidth]{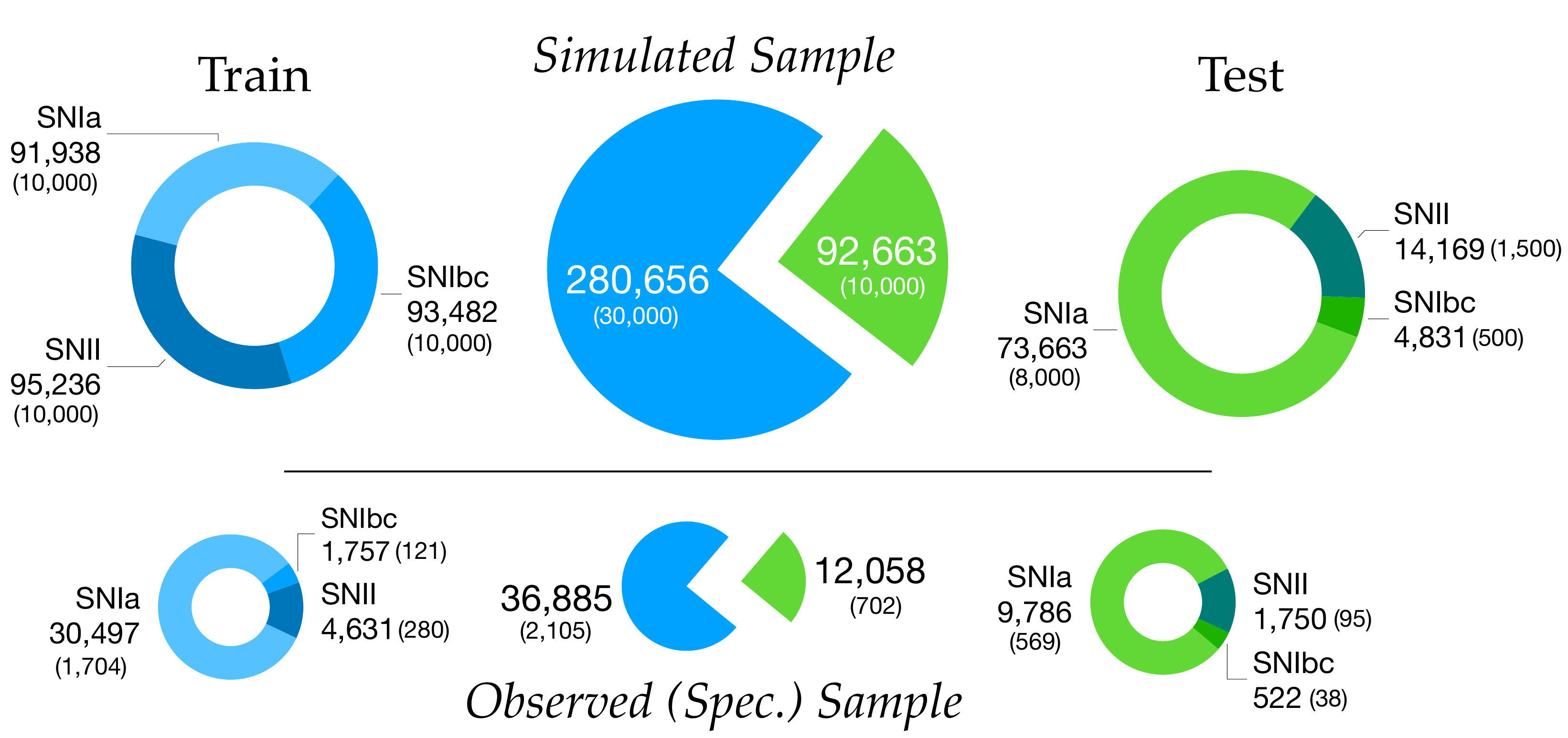}
    \caption{The training and testing datasets used in this work. \textbf{Above:} The composition of simulated events used in phase 1 of training the network. The total number of light curve segments in each sample are listed, and the number of unique transient events in each sample are given in parentheses. \textbf{Below:} The composition of spectroscopic events \textcolor{black}{from the ZTF Bright Transient Survey} used in the re-training stage (see text for details).}
    \label{fig:traintest}
\end{figure*}

\subsection{Phase Jitter of Synthetic Light Curves}
In our model, the date at which ZTF `discovers' an SN (deemed the trigger date, or $T_{\rm{trigger}}$) is the epoch when a second observation is taken with $S/N\geq5$. Because the date at which a transient is discovered is determined by the combination of a transient's brightness evolution and the survey cadence, this date should not be assumed to be either the date of the start of the explosion or a constant offset relative to it. Nevertheless, the simplicity of SN simulations may lead a machine learning model to over-train on unrealistic correlations in the trigger dates of the training set instead of the properties of the light curve. This will limit its use when applied to data collected from a survey with more restrictive, or relaxed, trigger criteria. To mitigate this issue, we remove observations from the first $d$~days of each transient light curve, where $d$ is a number randomly sampled from a truncated Normal distribution with $d\sim\mathcal{N}(T_{\rm{trigger}},\,0.25^{2})$ and $d\geq T_{\rm{trigger}}$. We then re-define the trigger date of each event as the epoch of the first observation in this new truncated light curve. For each light curve in both the simulated and spectroscopic sample, we calculate the phase of each observation relative to trigger in days. We correct these phases for time dilation using the photometric redshifts calculated above and correct light curve flux values for Galactic extinction. We then remove observations from each light curve obtained greater than 30 days after this new trigger date (as we are interested in this work in early classification). We do not k-correct our photometry, as this would encode assumptions about the shape of the SED of each transient class (and further encode the reliance on accurately determined redshifts).

Our classifier consists of a recurrent neural network, and these networks were initially constructed to process arrays of uniform length. To prepare irregularly-sampled light curves across multiple passbands for processing, common approaches include padding light curves to a fixed length and masking zero-valued observations in training \citep[e.g.,][]{2017Charnock_RNNPadding}; or interpolating observations onto a grid of uniform length \citep[e.g.,][]{2020Villar_superRAENN}.  We adopt both approaches, using a Gaussian Process model described in the following section.

\subsection{Gaussian Process Interpolation and Padding}
We use Gaussian Process Regression \citep[GPR;][]{2006Rasmussen_gaussianprocesses} to construct a light curve model for each transient event and interpolate observations in apparent magnitudes onto an evenly-spaced grid in time. In GPR, observations are considered to be realizations of a latent function with Gaussian noise. A function describing the covariance between observations, called the kernel, is constructed and its parameters are chosen to minimize the loss function (and maximize the likelihood of the obtained observations). This results in a continuous posterior distribution for a class of models that describe the data. GPR can additionally consider a mean model for the observations, and this further conditions the subsequent model predictions.

Our model, which is implemented in the lightweight code \texttt{tinygp}\footnote{\href{https://tinygp.readthedocs.io/en/stable/}{https://tinygp.readthedocs.io/en/stable/}}, uses a Mat\'ern kernel with $\nu=3/2$ that quantifies the correlation in time between observations $t_i$ and $t_j$ as 
\begin{equation}
    k(t_i, t_j) = (1 + \sqrt{3}r)e^{-\sqrt{3}r}
\end{equation}
In the above equation, $r = || \frac{t_i - t_j}{l}||_1$ parameterizes the time between observations and the scale factor $l$ is a free parameter. In addition to modeling the correlation between single-band observations in time, we model the correlations between \textit{bands}. We construct a $p \times p$ symmetric correlation matrix, where in this case $p=2$ is the number of passbands considered. The diagonal and off-diagonal terms (2 and 1 terms, respectively) are free parameters to be fit. 

We use a constant value in each band as our mean model ($\bar{g}$, $\bar{r}$), and add the uncertainty in each band to the diagonal of the covariance matrix along with additional free terms $j_g$, $j_r$ to capture any remaining measurement error. In total, our model consists of 8 free parameters. We transform our data from flux to magnitude space and minimize the negative marginal log probability of the model (our loss function) to determine the best-fit parameters for each event. We optimize our Gaussian Process fit with the \texttt{Scipy} package.

There is precedent to a Gaussian Process (GP) light curve model parameterized by the correlation in both wavelength/frequency and time. A similar method was first introduced by \cite{2019Boone_Avocado} with the photometric classifier \texttt{Avocado}, the winning submission of the Photometric LSST Astronomical Time-Series Classification Challenge \citep[PLAsTiCC;][]{2019Kessler_PLAstiCC}. In \texttt{Avocado}, the central wavelength of each passband was used as the coordinates of the wavelength dimension of the GP. \cite{2020Villar_AnomalyDetection} also used a custom GP model described by a kernel in both wavelength and time, where the distance metric in wavelength space was the Wasserstein-1 distance between each filter’s normalized throughput. Because these two GP methods place physical constraints on the covariance in wavelength, they are specialized cases of our model. A physically-motivated construction of the GP covariance matrix may be desirable when photometry is sparse; however, in this work we opt instead for a model with greater flexibility to fully characterize both the long-duration and short-duration behavior of each transient.

\begin{figure*}
    \centering
    \includegraphics[width=\linewidth]{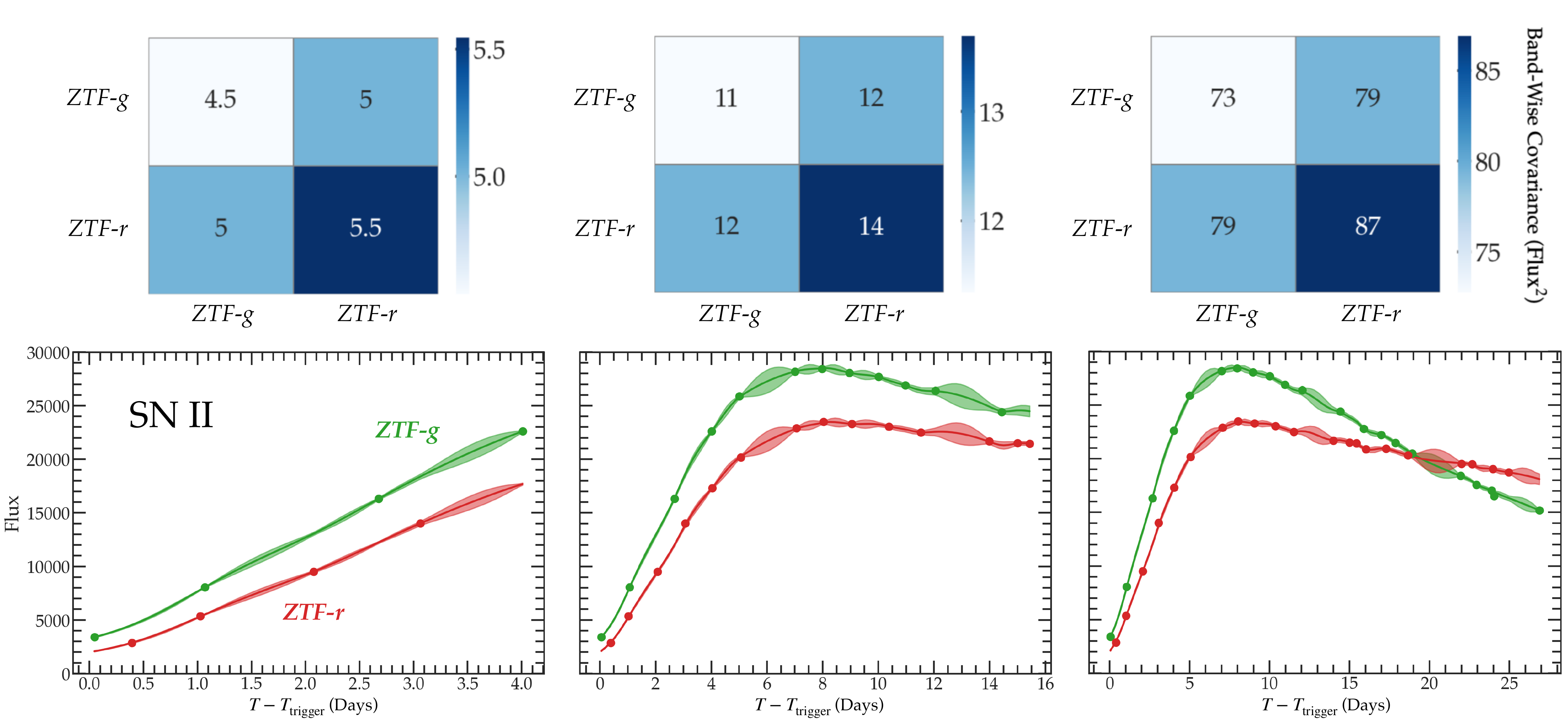}
    \caption{Best-fit Gaussian Process model for one simulated light curve in ZTF passbands (\textcolor{black}{ZTF-$g$} in green and \textcolor{black}{ZTF-$r$} in red) at three phases in its evolution from the time of trigger (bottom). The best-fit band-wise covariance matrix associated with each fit is shown at top. Because our model is parameterized by the covariance between observations in both time and wavelength, observations taken in a single band inform the model in the other.}
    \label{fig:GP}
\end{figure*}

We show the best-fit GP model for one simulated transient event across three phases of its light curve, in additional to the best-fit passband kernel at that phase, in Fig.~\ref{fig:GP}. With few observations, minimal correlation is detected between passbands (left plot). As more data is obtained (middle and right plots), the correlation increases and is used to improve the posterior light curve fits across both passbands. This learned correlation also ensures that the model in each band is robust against low-S/N observations (not shown). % \textcolor{red}{elnel now that we're only considering ZTF $g$ and $r$? I find these informative, but they probably wouldn't be useful to show with just two bands.}

%In addition to providing reasonable fits to the observed data in each band, the model is able to use data from one band to improve the fits in other bands where no data has been taken.

Although a GP model whose hyperparameters are optimized based on a full-phase light curve would best characterize an event's evolution in time and across passbands, this information will not be available in real-time. We decompose each transient light curve into segments to realistically simulate constructing a model at different phases of observation. From a light curve with $N$ total observations in the simulated sample, we construct a series of $N/2$ segments from the first $n$ observations, where $n=2,4...N$. In other words, we generate a unique segment every two observations of the light curve. This cadence was chosen to reduce the volumes of training data without decreasing the number of transient events they contain (and the associated photometric diversity). 

Multiple strategies have been proposed to pre-process unevenly-sampled and multi-variate time-series datasets for use in recurrent neural networks, and it remains unclear which strategy yields the best performance for SN classification. With this in mind, we construct three different data representations of our light segments:  
\begin{enumerate}
    \item \textbf{Raw and Pad Model:} In the first approach, we construct an array directly from the light curve observations. This photometry has been corrected for distance, extinction, and time dilation but has not been interpolated using our GP model. In this method, our temporal array contains the \textcolor{black}{(jittered)} phases at which observations were obtained in any band; the magnitude and magnitude uncertainty arrays are filled with zeros where a given band was unobserved. These arrays are then padded with zeros at the tail end so that the matrices for all events have equal length. 
    \item \textbf{100-Timestep Model:} In the second approach, we construct a GP fit for each light curve segment and compute the mean and standard deviation of the GP posterior distribution at 100 independently-spaced phases spanning the full extent of the segment. We then convert these interpolated observations and their uncertainties from magnitude space back to flux space. We note that, in contrast to prior light curve methods, this interpolation scheme does not maintain uniform spacing in time across light curve segments. One partial-phase segment may \textcolor{black}{consist of 10 observations spanning 5 days, in which case it will be interpolated onto a uniform grid with $\delta t = 0.05$ days; if 10 additional observations span another 5 days, the full 10-day segment would be interpolated onto a grid with $\delta t = 0.1$ days.}
    \item \textbf{0.2-Day Spacing Model:} In the third approach, we construct a GP fit as above and evaluate the GP model at phases beginning at the time of trigger and occurring every $\delta t = 0.2$ days until the end of the segment. We then convert these interpolated observations to flux. While this approach preserves consistent spacing in time, the interpolated light curves will not be equal in length. As in method 1, we pad each light curve array to a consistent length after interpolation. 
\end{enumerate}

We have adopted a flexible GP model over a physical light curve model to capture the observed diversity in SN photometry, which allows unexpected phenomenology to be expressed in the interpolated data. When an SN is young, however, this flexibility can be detrimental: with very few observations, the best-fit GP model may not reflect expected early-time SN behavior. To bypass this limitation, we only interpolate our observations after 5 or more total observations have been obtained. Before this phase, we pad the photometry with zeros to a uniform length in all three of our models and pass these arrays directly into the neural network \textcolor{black}{(as with our `Rad and Pad' model above)}. We note that a more sophisticated approach may be to integrate a physically-motivated light curve model at early times, then transition to a more flexible GP model as more data is obtained. We leave this implementation to future work.

After obtaining the interpolated or padded representations of each light curve segment, we convert its flux to normalized luminosity using the equation
\begin{equation}
    L_{\rm{Norm}} = \frac{4 \pi d^2 F}{\alpha}
\end{equation}
where $F$ is the GP-interpolated flux in each band; $d$ is the luminosity distance to the transient in parsec, estimated from the photometric redshift of the host galaxy; and $\alpha=10^{23}$ is a normalization constant used to keep the computed luminosities near unity (as neural network performance is often better if the input range is constrained). Photometric uncertainties extracted from the GP fit are similarly converted as $\sigma_{\rm{L, Norm}} = L_{\rm{Norm}} \sigma_{F}/F$. 

\textcolor{black}{Each interpolated or padded transient light curve segment is represented in our dataset as an \textcolor{black}{$N_{\rm{pad}} \times p$} matrix, where $N_{\rm{pad}}$ is the \textcolor{black}{full length of each light curve segment, including padded values} (this value is different for each of the three methods) and $p=2$ is the number of passbands (ZTF-$g,r$). To this matrix, we add the array of interpolated phases $t = T-T_{\textrm{trigger}}$; the GP-derived photometric uncertainties in each band; and \textcolor{black}{$N_{\rm{pad}}$}-length arrays contains the fixed repeated values of \newline $E(B-V)_{MW}$, $z_{\rm phot}$, and host galaxy normalized luminosity in \textcolor{black}{PS1-$grizy$} filters. This results in a data matrix containing arrays of dimensionality \textcolor{black}{$N_{\rm{pad}} \times (2p + 8)$} that \textcolor{black}{are} used to both train and validate the network.}

\textcolor{black}{We consolidate our \textcolor{black}{$N_{\rm{pad}} \times (2p + 8)$} arrays for all SN segments associated with the SNe in our training samples described in \ref{subsec:sim_lcs}; this forms our training set. The segment arrays associated with the test SNe similarly form our test set. Our target array for prediction is a one-hot encoded representation of our three classes: SNe~II are encoded as `0', SNe~Ia as `1', and SNe~Ib/c as `2'. We report the number of unique transient events across the simulated and spectroscopic training and test set, along with the number of light curve segments in each sample, in Fig.~\ref{fig:traintest}.}

\section{Model Architecture}\label{sec:architecture}
\subsection{Overview}
\label{subsec:architecture_overview}
% may vary slightly (as the number of 
% Fig.~\ref{fig:traintest}. 
\textcolor{black}{To take advantage of the temporal correlations embedded within our partial-phase time-series samples, we design a recurrent neural network for photometric classification.}

A recurrent neural network (RNN) is comprised of multiple connected and directed ``units'' that are trained to learn a data representation from both the internal states of earlier units as well as a component of the input array. Because gradients propagate through the network, RNNs have traditionally suffered from vanishing and exploding gradients (those tending toward 0 and infinity, respectively) in training. This issue can be alleviated by the use of recurrent units that modulate information flow through the network; for example, Gated Recurrent Units \citep[GRUs;][]{cho-etal-2014-learning} feature coupled `update' and `reset' gates.  We have selected the more conventional Long-Short Term Memory cells \citep[LSTM;][]{hochreiter1997_LSTM} as our RNN units. LSTMs contain three gates: a `forget' gate that removes information from the cell state; an `input' gate that regulates the adding of new information from the data sequence into the cell state; and an `output', which regulates the output of the cell state into a subsequent unit. We use LSTMs for their ability to retain insights from the long-term behavior of complex time-series data, where they often achieve superior performance to GRUs \citep[e.g., ][]{2021Cahuantzi_LSTM}. %Despite consisting of more parameters than GRUs and requiring more data to train, w

\begin{figure}
    \centering
    \includegraphics[width=\linewidth]{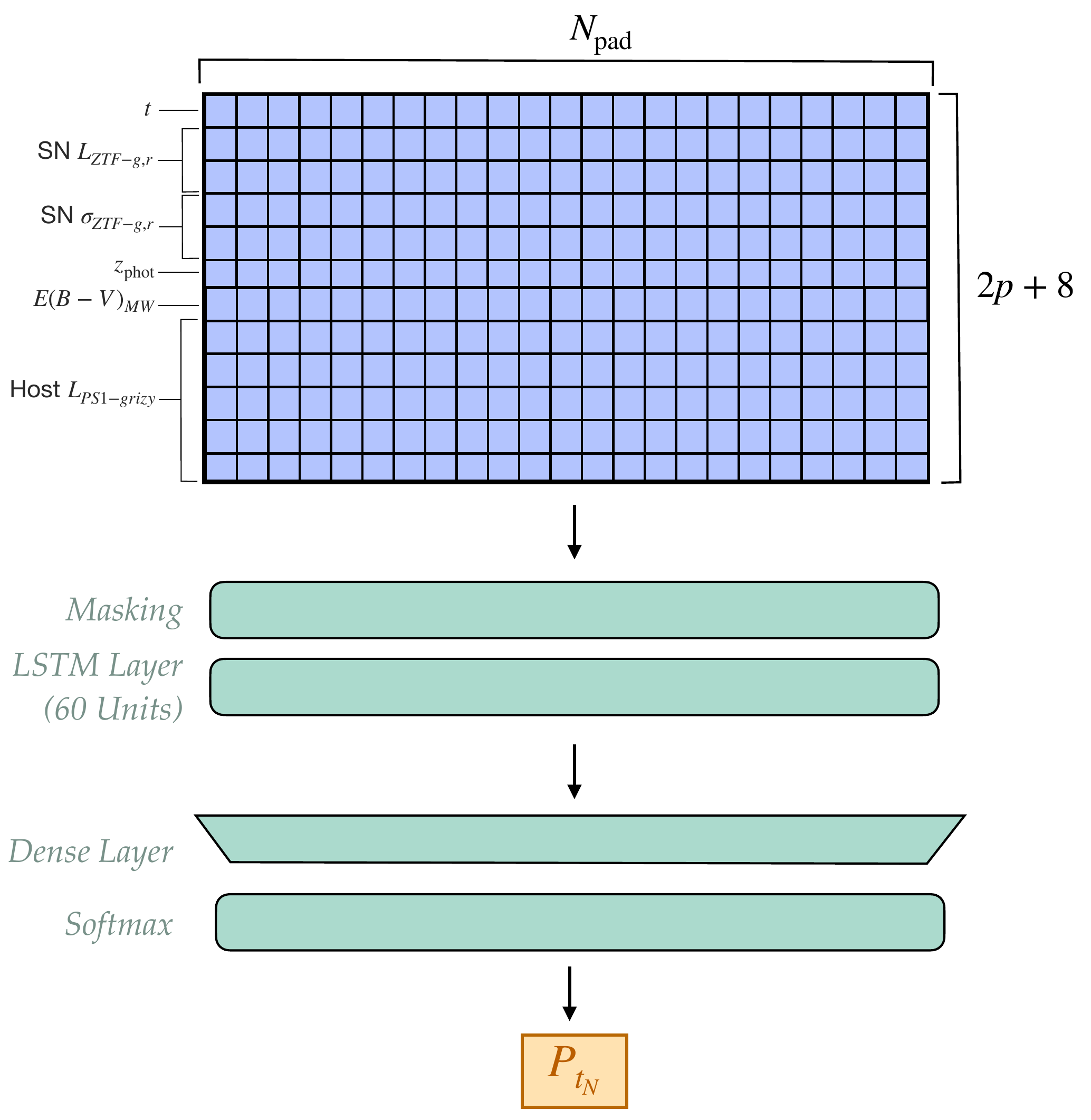}
    \caption{Architecture for our RNN, which is composed of a single LSTM layer of 60 units. The photometry in each band (interpolated, padded, or both) is stacked with the photometric redshift, extinction, and host galaxy photometry, which are repeated at each phase. This combined array serves as the input to our RNN, and the network returns a single classification prediction corresponding to the light curve segment processed.}
    \label{fig:rnn_architecture}
\end{figure}

The architecture of our neural network is illustrated in Fig.~\ref{fig:rnn_architecture}. It consists of a single LSTM layer of 60 units, with a \texttt{sigmoid} activation function defined as 
\begin{equation}
    S(x) = \frac{1}{1 + e^{-x}}
\end{equation}
where $x$ is the input tensor to the gate. The next layer is a masking layer, which constructs a boolean mask of equal dimensionality to the input data. \textcolor{black}{This mask, which is propagated through successive layers along with the data, is constructed to flag and ignore padded values in training.} This layer is followed by a dense layer to consolidate model outputs into a set of normalized probability scores corresponding to the likelihood of the light curve to belong to the three transient classes considered (SN~Ia, SN~II, SN~Ib/c).

\textcolor{black}{Many previous RNN architectures for photometric classification have applied dropout layers that deactivate randomly-selected neurons during training to ensure that all neurons are used; batch normalization is also common to decrease the number of training epochs needed to achieve accurate classification results. We have found that neither modification significantly increases the accuracy of our model when applied to real data, and so have excluded them. Our network is implemented in \texttt{TensorFlow} \citep{tensorflow2015-whitepaper} using the \texttt{Keras} \citep{chollet2015_keras} interface.}

\subsection{Class-Weighted Loss For Accurate Early Predictions}\label{weightedLoss}
A common choice for the loss function used to train multi-class classification networks is the sparse categorical cross-entropy. For an event belong to true class $a$, the categorical cross-entropy is calculated as
\begin{equation}
    L(y, \hat{y}) = -\sum_{j=0}^M \sum_{i=0}^C y_{ij} \times \textrm{log}\left( \hat{y}_{ij} \right)
\end{equation}
where $C$ is the number of transient classes (in this work $C=3$), $M$ is the number of events in a training batch, $y_{ij}$ is the true class of event $j$ after one-hot-encoding (with $y_{ij} = 1$ where $j=a$, else $y_{ij} = 0$), and $\hat{y}_{ij}$ is the set of predicted normalized probabilities output from the softmax layer of the network.
The cross-entropy for all events in a given batch are summed at each training epoch to compute the network loss. 

We modify this framework to weight the contributions of particular segments over others. The modified cross-entropy loss is constructed as:
\begin{equation}
    L(y, \hat{y}) = -\sum_{j=0}^M  \sum_{i=0}^C \beta_{ij} y_{ij} \times \textrm{log}\left( \hat{y}_{ij} \right)
\end{equation}
We calculate the weight of each light curve segment as a function of $t_N$, the final phase covered by the segment \textcolor{black}{(ignoring padded values)}:
\begin{eqnarray}
\beta_{ij}(t_N) =
  \begin{cases}
    10, & \text{for } t_N \leq 3 \text{ days} \\
    5, & \text{for } t_N>3  \text{ days} \text{ and } t_N \leq 15  \text{ days} \\
    1, & \text{otherwise}\
  \end{cases}
\end{eqnarray}
These weights are constructed so that the network minimizes the loss by prioritizing the classification of SNe using only photometry provided in the first three days following detection. In the re-training stage using observed ZTF light curves (see \textsection \ref{sec:training} for details), we modulate these values by an additional weight according to the class of the transient associated with the segment:
\begin{eqnarray}
\beta_{ij}(t_N) =
  \begin{cases}
    \beta_{ij}(t_N)\times 2, & \text{for } j=0 \\
    \beta_{ij}(t_N)\times 1, & \text{for } j=1 \\
    \beta_{ij}(t_N)\times 5, & \text{for } j=2 \\  \end{cases}
\end{eqnarray}
The multiplicative factors are added to make the network more sensitive to identifying SNe~II and SNe~Ib/c, which are poorly represented in our re-training dataset. SNe~II events are more photometrically heterogeneous than SNe~Ia, and SNe~Ib/c are more easily confused with SNe~Ia; the network learns to classify these more challenging minority classes only through explicit weighting. We note that we have tested the performance of the network after using these class weights in the primary training stage (on synthetic data before re-training). We find no significant improvement in performance relative to the early-weighted cross-entropy loss. The same principle we have applied here can be used to adapt publicly-available classifiers for specific science cases (e.g., to prioritize SN~Ia classification for cosmological studies, or to prioritize late classification for nebular-phase spectroscopic follow-up).  %(this challenge is also noted in \citealt{2022Aleo_YSEDR1}). 

\section{Primary and Adaptive Training}\label{sec:training}
Because of the inherent complexity of observational data, classification models trained on simulated data typically under-perform on more realistic datasets. We have attempted to reduce the discrepancy between synthetic and observed samples by simulating \textcolor{black}{realistic} photometric redshifts and jittering the early photometry of simulated light curves. We further prepare our trained network for real-time operation with a domain-adaptation step in the training process. There are two primary motivations for this:
\begin{itemize}
\item \textbf{Encoding a weak dependence on class imbalance and survey systematics:} If the network was fully trained using realistic relative class rates, it may prioritize accurate classification of some classes over others (e.g., only classifying SNe~Ia, the most common SN class in observed samples); conversely, a network trained on evenly balanced datasets may predict unrealistically-large numbers of minority classes. A re-training stage allows a reasonably flexible architecture to consider both pieces of information. After learning the underlying phenomenological differences between the light curves of each SN class from the balanced simulated sample, the network in its re-training stage additionally considers the observed distribution of real events in a given survey. 
%\item \textit{Transferring  Between Photometric Systems:} The photometry of host galaxies simulated in SCOTCH \citep{2022Lokken_SCOTCH} was reported in Vera Rubin Obs. passbands. While this is valuable for considering the upcoming performance of photometric classifiers using Rubin data, we aim to validate the performance of our network \textit{now} on observed samples. Our two-stage model training allows the network to first consider the correlations between transient type and their host galaxy photometry, and then to optimize its performance to the PS1 filter system.
\item \textbf{Increasing the Realism of Estimated Photo-$z$s:} Given the reliance of photometric classifiers on the predicted distance to the transient, it is imperative that our model is trained on a set of photometric redshifts with statistical properties comparable to those derived from observed photometry. \textcolor{black}{While the overall scatter of the simulated and observed samples appears similar, re-training prevents the network from relying upon unphysical artifacts within the simulation for its prediction.}
\end{itemize}

First, we train our network on the simulated training sample for 200 epochs. We use the stochastic gradient descent optimizer \texttt{Adam} \citep{2014Kingma_Adam}, with a learning rate of $l=10^{-4}$. We then re-train this model for 200 additional epochs and the same learning rate on the observed training sample. The temporal weights used in both training stages are identical; however, we have only used the class weights in the re-training stage, to account for the class imbalance. We use a training batch size of \textcolor{black}{32} in the simulated training stage and 12 in the spectroscopic training stage to account for the smaller data volumes of the latter dataset. We conducted the primary and secondary training for each of the three models across 24 CPUs of a node of the Flatiron Institute Scientific Computing Hub. The primary training stage was completed in a wall time of 31.2 - 32.1 hours per model, while the secondary training stage was conducted in 9.7 - 11.6 hours of wall time per model. 

\section{Results}\label{sec:results}
%We provide an example of the classification predictions from our raw-photometry model tested on multiple segments of a single transient from the ZTF BTS in Fig.~\ref{fig:SN2020uxz}. \textcolor{red}{Edit!}
In the following section, we evaluate the performance of our three models as a function of phase. We construct ROC curves and PRC curves for each model, and report balanced metrics across the SN classes in our sample. 

\subsection{\textcolor{black}{Receiver Operator Characteristic} and Precision-Recall Curves}\label{subsec:ROCCurve_RawPhot}
For both the simulated and the observed samples, we bin our light curve segments into early ($t_N < 3$ days), intermediate ($3 < t_N < 15$ days), and late ($15> t_N >30$ days) epochs depending on the phase of the final observation in the segment. The vast majority of events within the `early' bin will still be brightening, while the late events will predominantly be dimming after peak.  

For a given phase bin, events correctly classified as belonging to \textcolor{black}{a} class $a$ are called True Positives ($TP$). Events incorrectly classified as belonging to class $a$ are called False Positives ($FP$). Conversely, events correctly and incorrectly classified as \textit{not} belonging to class $a$ are deemed True Negatives ($TN$) and False Negatives ($FN$), respectively. Then, the classification precision, also known as its purity, is calculated as 
\begin{equation}
    p_a = \frac{TP_a}{TP_a+FP_a}
\end{equation}
The classification recall, also known as its completeness, is calculated by
\begin{equation}
    r_a = \frac{TP_a}{TP_a+FN_a}
\end{equation}
A common method to evaluate the performance of a binary classifier is to consider the false positive rate $FP_a$ as a function of the true positive rate $TP_a$; a well-trained classifier should be able to maximally increase the rate at which they recover events from class $a$ while minimizing the rate at which they mistakenly identify the alternative class as belonging to class $a$. The ideal behavior of this $TP-FP$ curve, also known as a Receiver Operator Characteristic Curve \citep[ROC;][]{peterson1954_ROCtheory}, is to rapidly increase toward a true positive rate of unity with minimal false positives. We quantify this behavior by calculating the Area Under the ROC \citep[AUROC;][]{pepe2006_AUROC}, which is unity in the limit of perfect classification and 0.5 in the case of random guessing. 

Another common metric is to plot the recall of the model as a function of its precision; a well-performing classifier exhibits a Precision-Recall (PR) curve with high recall rates (the fraction of recovered events from class $a$) even as the demanded model precision increases. As in the ROC curve, this trade-off can be quantified by the area under the precision-recall curve \citep[AUPRC;][]{saito2015_precision}, with AUPRC$\;=1$ for perfect classification and AUC$\;=f_a$, the fraction of the test data consisting of class $a$, for random guessing. The AUROC and the AUPRC provide complementary views of a classifier's performance. 

%$Because our work is motivated by the need to rapidly select SNe for follow-up, w precision of our transient classifier over its recall. 

Our work is motivated by the need for rapid characterization of common events to conduct follow-up with highly limited resources. We will be unable to study the vast majority of discovered transient events in detail, but we must ensure that we do not waste precious resources when we do. For this reason, this and other upcoming SN classifiers will need to optimize a method's precision over its recall. We note that if an event is intrinsically rare, a classifier's recall rate \textcolor{black}{may} be prioritized over its precision.

We use the \texttt{StratifiedKFold} module in \texttt{sklearn} to generate five evenly-sized random splits of the test datasets. These data subsets are mutually exclusive: no events are repeated between them. Next, we binarize our classification predictions for each split  using the one-versus-all approach (in which the true class $a$ is assigned a value of 1 and the other two classes are assigned a value of 0). We then generate ROC and PR Curves for each data split and each phase range (`early', `intermediate', and `late'). Finally, we calculate the mean and standard deviation of the curves across all splits.

\begin{figure*}
    \centering
    \includegraphics[width=0.8\linewidth]{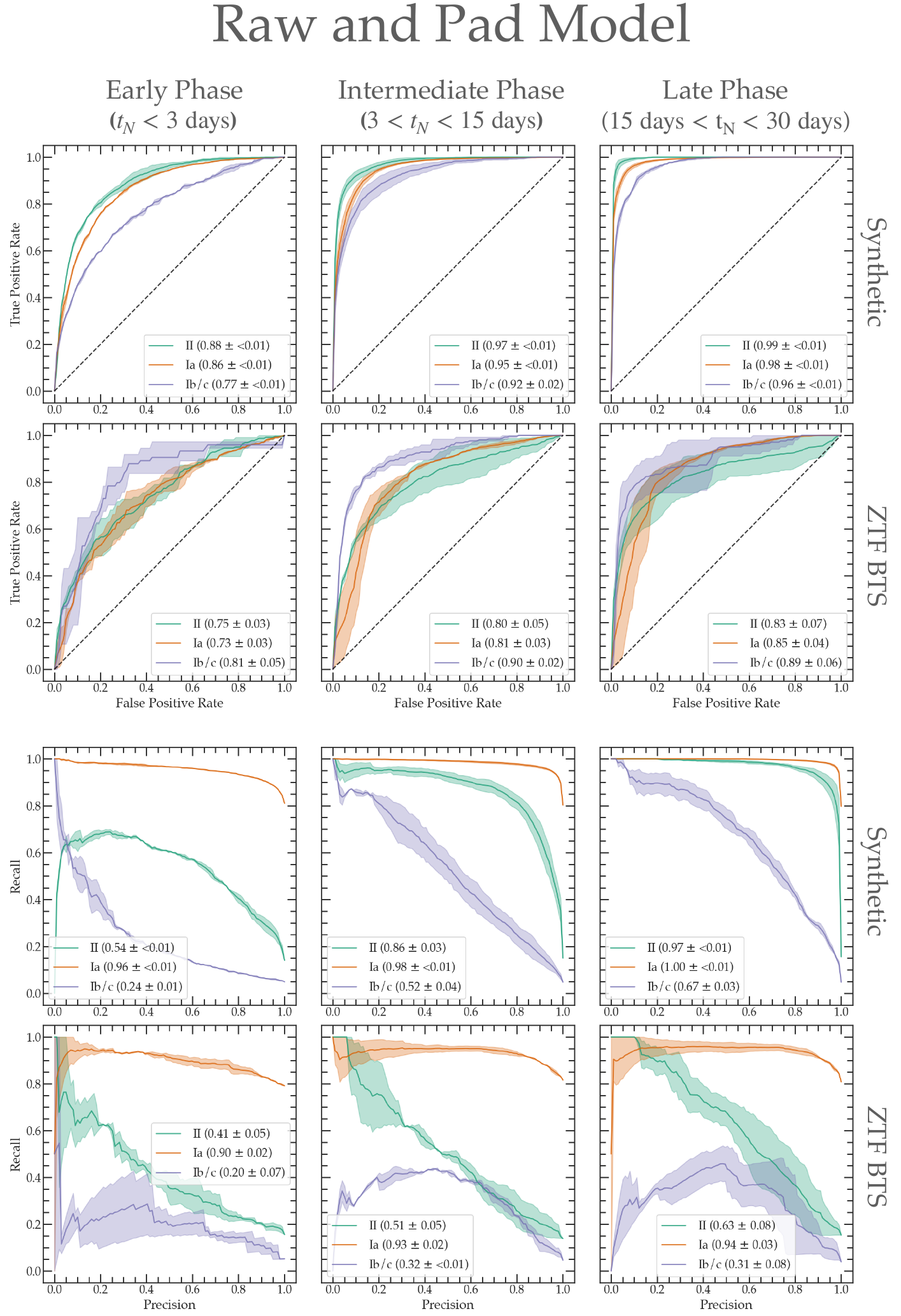}
    \caption{\textbf{First Row:} Receiver Operator Characteristic (ROC) Curves for the network trained and tested directly on transient and host photometry (without using a GP interpolation scheme). The performance has been evaluated on light curve segments truncated at early, intermediate, and late times. The mean AUROC are reported with 1-$\sigma$ uncertainties in the legend of each panel. \textbf{Third Row:} The Precision-Recall Curves for the same model applied to synthetic photometry. The AUPRC with 1-$\sigma$ uncertainties are given in the corresponding legends.  \textbf{Second and Fourth Rows:} ROC Curves, and Precision-Recall Curves, for the same model after 200 epochs of re-training on observed light curve segments (see text for details).}
    \label{fig:rawROC}
\end{figure*}

\begin{figure*}
    \centering
    \includegraphics[width=0.8\linewidth]{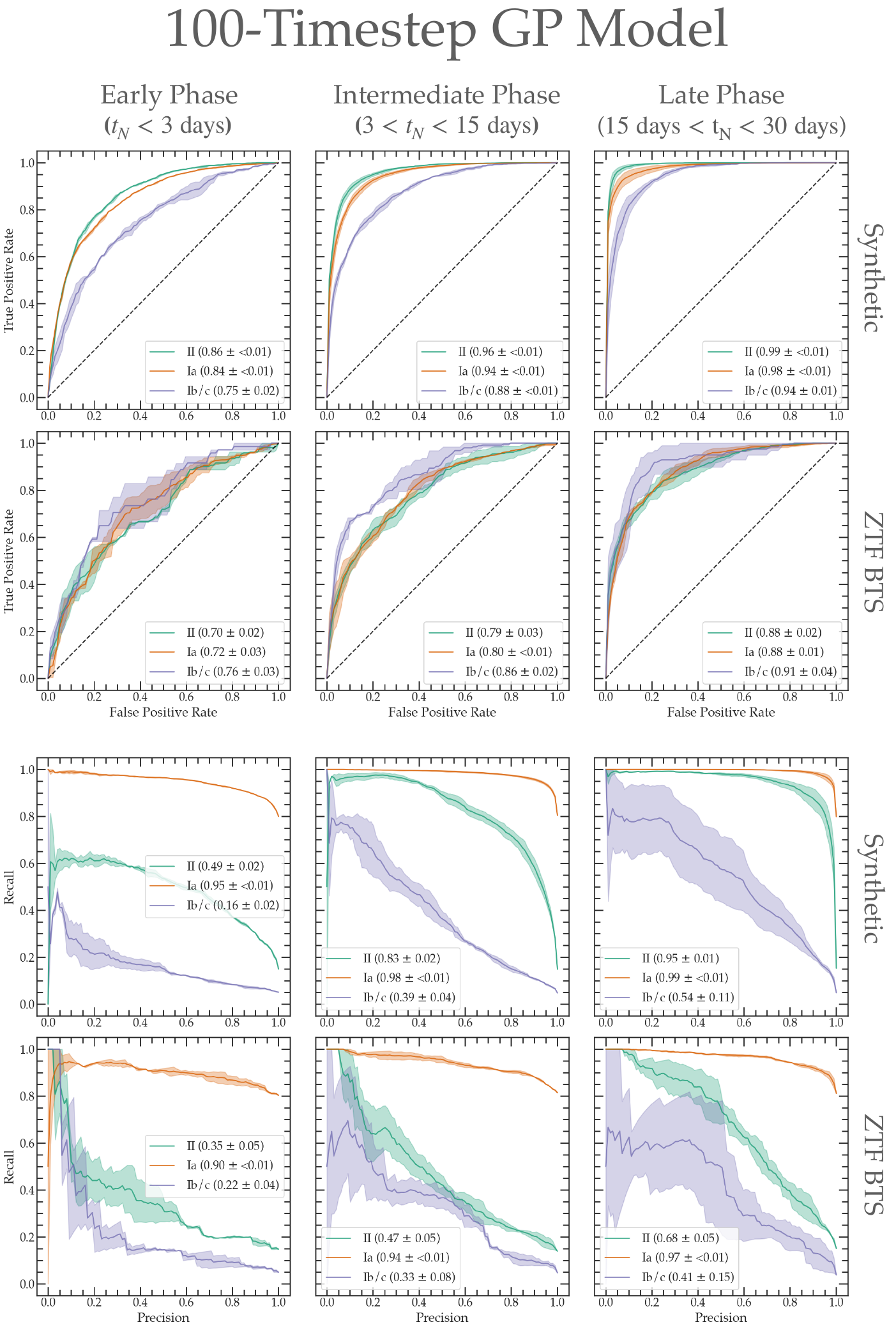}
    \caption{Same plot as Fig.~\ref{fig:rawROC} for the 100-Timestep GP model.}
    \label{fig:GP100ROC}
\end{figure*}

\begin{figure*}
    \centering
    \includegraphics[width=0.8\linewidth]{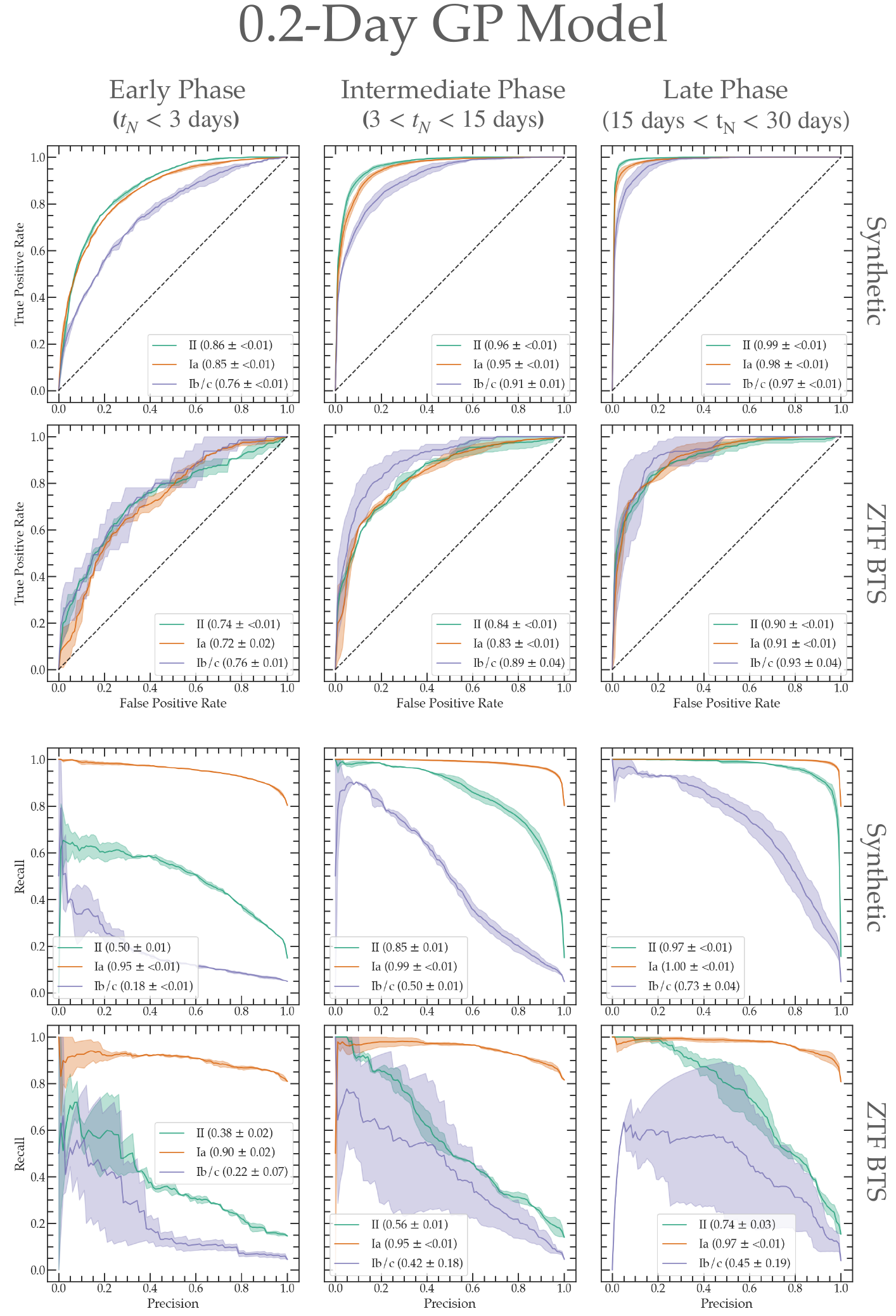}
    \caption{Same plot as Fig.~\ref{fig:rawROC} for the  0.2-Day  GP model.}
    \label{fig:GP0pt2ROC}
\end{figure*}

The ROC and PR curves for the `Raw with Pad', `100-Timestep GP', and `0.2-Day GP' models are shown in Figures \ref{fig:rawROC}, \ref{fig:GP100ROC}, and \ref{fig:GP0pt2ROC}, respectively. We show these curves applied to both the simulated test segments after the first training stage and on the observed test segments after the retraining stage. We report the class-specific AUC and AUPRC values in the figure legends. 

On the simulated sample, we achieve near-perfect classification for both SNe~II and SNe~Ia at late phases ($15<t_N<30$): Our \textcolor{black}{mean} AUROC values for the two classes are $\geq98$\% across all three models\textcolor{black}{, and the mean AUPRC values are $\geq95$\%}. We observe \textcolor{black}{substantially} worse results at late phases for SNe~Ib/c: \textcolor{black}{in the worst case, we observe an} AUROC of \textcolor{black}{$94\pm1$\%} for the 100-Timestep GP model but an associated AUPRC of \textcolor{black}{$54\pm 11$\%}. We attribute this to the class imbalance of the test set, to which the the AUPRC is particularly sensitive: $\sim5\%$ of the sample is SN~Ib/c. 

At \textcolor{black}{late} phases, we \textcolor{black}{also} observe \textcolor{black}{the largest} difference between the three classifiers tested on the simulated transients. \textcolor{black}{Despite generally comparable performance for all models, we find the lowest metrics} with the 100-Timestep GP model, with AUPRC values of $99\ \pm<1$\%, $95\pm1$\%, and $54\pm11$\% for SNe~Ia, SNe~II, and SNe~Ib/c, respectively. It is possible that this is a reflection of the loss of physically-relevant timescales in pre-processing these light curves: because 100 observations are generated independent of the duration of the light curve segment, the same rise \textcolor{black}{and decline times} could comprise half of one processed array and span the full extent of another. Although we also pass the phase information explicitly to the network, we propose that the physically-motivated spacing between array elements aids the network in learning the temporal evolution of an event. Further, we note that the first five observations were passed in directly (without interpolation) for each of the models. Because ZTF observations are taken with a median cadence of $\sim$2 days, these observations encode physical information about the phase of the event in the number of array elements. The network trained to \textcolor{black}{leverage this information from raw data (the `Raw and Pad' model) is better-adapted to process this early data, evidenced by the highest early-phase precision for simulated SNe~Ib/c (24$\ \pm 1$\%).} %(the `Raw and Pad' and the `0.2-Day GP' models) are better-adapted to process the raw early data than the 100-Timestep model. 

%Finally, we note that the AUC and AUPRC values for each class increase monotonically from the early-phase to the late-phase bin. This aligns with our expectation that as more data is taken, our ability to classify an explosion improves. We conclude that the RNN is successfully able to improve its performance according to the temporal evolution of the simulated datasets, \textcolor{black}{regardless of the pre-processing method adopted.}

We now consider the performance of our three models on the spectroscopically-confirmed sample of SNe from the ZTF BTS. For the dominant class in the sample, SN~Ia, we note surprisingly consistent results when transitioning from simulations to observations: the late-phase classification achieves an AUROC of $85\pm 4$\% and an AUPRC of $94\pm3$\% in the worst case (with the `Raw and Pad' model). The performance on the other two classes is appreciably worse: we observe significantly greater variance among the performance results between the five cross-validation folds for each the three models, particularly for SNe~Ib/c. In the worst case, our mean AUROC metric for late-phase classification decreases by 36\% for SNe~Ib/c and 34\% for SNe~II with the `Raw and Pad' model. %When applied to the real data, we find the worst performance with this model: \textcolor{black}{.} %our early-phase AUPRC suggests SN~II classification is barely superior to random guessing ($20\pm7$\% compared to its 15\% abundance in the dataset).

\textcolor{black}{Our AUROC and AUPRC values increase roughly monotonically with phase for each classifier applied to both the simulated and observed SN samples. There is a single exception: the mean SN~Ib/c AUROC decreases from intermediate to late phase with the `Raw and Pad' model, although the difference is not statistically significant. The `0.2-Day GP' model achieves marginally superior performance at late-phase, and we conclude that this representation may be slightly better able to incorporate the long-term evolution of each light curve. We note that, in an earlier iteration of this work, the inclusion of ZTF SNe that did not pass the ZTF-imposed quality and purity cuts caused this performance increase with phase to disappear for the the `Raw and Pad' model. The fact that our simplified network architecture is able to reasonably fit the temporal behavior of the real data across all three light curve representations suggests that, for sufficient-quality observations, interpolation onto an evenly-spaced grid (as is commonplace in present-day photometric classifiers) may not be a required pre-processing step for accurate RNN classification.}%, we do not observe the same trend for all phases in the observed SN sample. Our highest SN~II AUROC is achieved only for intermediate-phase light curves ($81\pm <1$\% versus $78\pm <1$\% at late phase), and the difference between the AUPRC values of each class from intermediate to late-phase prediction is not statistically significant. The increase in model performance is more evident among the two GP models applied to the real data. This suggests that our simplified RNN architecture struggles to learn the temporal behavior of the real data without interpolation, emphasizing the need to fit observations to some underlying model as a pre-processing step for real-time classification.

\textcolor{black}{Finally, we find that \textit{all three models are able to achieve classification performance in the first three days better than random guessing across all SN classes.} For the `0.2-Day GP' model, we find mean AUROC values of $>72$\% across all classes, and an AUPRC value of 90\% for SNe Ia.} We additionally find AUPRC values of \textcolor{black}{$22\pm 7$\% for SNe~Ib/c and $38\pm 2$\% for SNe~II}. These are encouraging results that the light curve rise of an SN, coupled with its host galaxy photometry, can be used for real-time early classification.  Our results also suggest that the application of machine-learning solutions to simulated SN samples alone may suggest overly optimistic performance, and may not be \textcolor{black}{reflective} of their performance on real data.

\subsection{Comparison to Light Curve-Only Classification}\label{subsec:component_comparison}
\textcolor{black}{Next, we quantify the impact of each component of our classification framework on our final results. Due to the slightly superior late-time performance of the `0.2-Day GP' model described in the previous section, we use this as our baseline for comparison.}

\textcolor{black}{To facilitate this analysis, we have trained three additional models using the `0.2-Day GP' interpolation method. In the first, we remove host-galaxy photometry and consider only the transient's light curve, photometric redshift, and the Galactic extinction along the line of sight. In the second, we remove our primary training stage, and train our network exclusively on the imbalanced training set from the ZTF BTS. In the third, we remove our adaptive training stage and train our network exclusively on the balanced simulated sample.}

To better facilitate a comparison \textcolor{black}{between these models}, we consider the macro-average of the AUROC, AUPRC, precision, and recall between the three classes. These metrics consider the contributions of each class equally, and consequently are not dominated by the network's performance on the most populous class in the observed sample (SN~Ia). This is in contrast to the micro-average, in which a weighted average is computed using the number of events of each class. \textcolor{black}{Our macro-averaged} balanced metrics are denoted with a `b-' prefix. We also introduce an additional metric to our comparison: the balanced-F$_1$ score, defined as the harmonic mean between the precision and recall of a classifier averaged between classes:
\begin{equation}
    F_1 = \frac{1}{C} \sum_{i=0}^C \frac{2 r_i  p_i}{r_i + p_i}
\end{equation}

The balanced-$F_1$ score consolidates multiple performance metrics into a single value for directly comparing classifiers. \textcolor{black}{We report this and our other metrics for the above models evaluated on the early-phase ($<3$~day) light curves of the observed ZTF BTS sample. We report our results in Table~\ref{tbl:host_and_retrain}.}

\textcolor{black}{We find that \textit{removing any component of our training sequence decreases the mean value of every metric}. The largest decrease in every metric comes from removing the adaptive training stage. Interestingly, removing the primary training stage results in a nearly comparable mean sample accuracy to our baseline classifier (81\% compared to 82\%). Removing the adaptive training stage results in a substantial drop in mean accuracy (66\% compared to 82\%), although these two truncated-training networks have comparable precision. This can be understood in the context of the class breakdown in each training set: Training only on the fully balanced simulated sample results in inaccurate predictions on a sample comprised primarily of SNe~Ia, while the opposite holds true for a training set dominated by SNe~Ia. The smallest decrease comes from the removal of host-galaxy information. Although the uncertainties are large, the systematic decrease in every property suggests the benefit of including this photometry in early classification efforts.}

\subsection{Comparison to Previous Classifiers}\label{subsec:results_comparison}
We \textcolor{black}{now} compare our classifier to pre-existing methods in its \textcolor{black}{design}. Because we have sought to overcome existing barriers to real-time classification using physically-motivated information instead of model complexity, our classifier is simpler in architecture to the majority of comparative leading methods. Earlier RNN methods for photometric classification (e.g., \texttt{superNNova}; \citealt{moller2020supernnova}) had considered two layers of bi-directional GRUs, in which the RNN representation of the light curve at each unit is informed by information from both earlier and later phases. These architectures are suited for full-phase classification but are ill-suited for classification with an incomplete light curve. 

\texttt{RAPID}, the first recurrent neural network to be constructed for real-time classification, consisted of two layers of 100 uni-directional GRUs to avoid leaking information from later phases. It was later updated to use Temporal-Convolutional Neural units \citep[TCNs;][]{2018Bai_TCN}. Our method uses a single LSTM layer consisting of 60 uni-directional units. To interpolate observations onto a uniform grid, \cite{muthukrishna2019rapid} interpolated synthetic ZTF observations onto a grid of 50 observations spanning $-70<T - T_{\rm{trigger}}<80$ days at a 3-day cadence. Model predictions were then wrapped in a \texttt{TimeDistributed} layer to provide a classification output at each interval in phase. In contrast, we process segments of the light curve separately through our network and interpolate each segment \textcolor{black}{independently} with a flexible GP. We have chosen this approach so that our light curve representation improves as additional observations are obtained, contrary to the stationary interpolation scheme considered by \cite{muthukrishna2019rapid}. This mimics model construction in real-time: as additional observations are obtained, our understanding of the shape and evolution of the light curve improves. \cite{muthukrishna2019rapid} also define a pseudo-class to distinguish pre-explosion from post-explosion phases, and use a parametric fit to each light curve in training to estimate the time of explosion relative to trigger. We avoid this framework so that the network does not need to estimate the time of explosion, which in the case of sparse light curves is non-trivial and non-essential for classification. We note that our model can easily be modified to incorporate pre-explosion non-detections, as is done in \citep{muthukrishna2019rapid}; this may further improve its performance.

\textcolor{black}{Two years after the development of \texttt{RAPID}, t}he real-time classifier \texttt{SCONE} \citep{2021SCONE} was developed. \texttt{SCONE} uses a convolutional neural network applied to 2-dimensional `flux-heatmap' representations of transient light curves for classification, after GP interpolation in wavelength and time. The model consists of 22,606 free parameters for multi-class classification, whereas ours consists of 17,463. We note, however, that our model has only been constructed to distinguish three classes of SNe, whereas theirs includes SNe~Iax, SNe~91bg, and SLSNe-I for a total of six possible SN classes.

During the writing of this work, \cite{2022Pimentel_DeepSNClassification} released a Deep Attention Model for photometric classification of transients discovered by ZTF. Attention Models \citep{2014Mnih_Attention} are secondary neural networks added to a primary network and tasked with learning the input features most relevant to the output of the primary network. Using a custom time-modulated attention component (called `TimeModAttn'), \cite{2022Pimentel_DeepSNClassification} uses an auto-encoder to learn a continuous representation of a light curve from discrete observations. Two model approaches are considered: a serial one in which an auto-encoder simultaneously encodes the light curves in ZTF-$g$ and ZTF-$r$ into a single latent vector; and a parallel one in which the light curves of an event in each band are encoded separately, combined, and then compressed with a linear projection to arrive at a dimensionality comparable to that of the serial model. After the partial light curve is encoded, it is processed by a Multi-Layered Perceptron (MLP; \citealt{MURTAGH_MLP}) model with two layers, a dropout fraction of 50\%, a batch normalization component, and a final softmax layer to convert the output of the model to a series of classification probabilities. We note that this requires more pre-processing than our model, and we require no dropout, batch normalization, or attention component. Nevertheless, the inclusion of the attention component allows for the network to predict the specific observations most \textcolor{black}{relevant} for classification by the model, a major advancement in the interpretability of machine learning models used for \textcolor{black}{photometric classification}.

\begin{center}
\begin{table*}[!htp] %\hspace*{-12mm}
\begin{tabular}{lllllll}\hline

Model  & b-AUROC  & b-AUPRC & b-Precision & b-Recall  & b-F$_1$ Score & Accuracy  \\ \hline \hline

Baseline  
& 0.74 $\pm$ 0.04  %b-auroc
& 0.52 $\pm$ 0.07 %b-auprc
& 0.58 $\pm$ 0.13 %b-precision
& 0.46 $\pm$ 0.09 %b-recall
& 0.48 $\pm$ 0.11 %b-F1
& 0.82 $\pm$ 0.02  %accuracy 
\\ \cline{2-7}

No Host  
& 0.72 $\pm$ 0.08 %b-auroc
& 0.48 $\pm$ 0.09 %b-auprc
& 0.48 $\pm$ 0.12 %b-precision
& 0.41 $\pm$ 0.09 %b-recall
& 0.40 $\pm$ 0.08 %b-F1
& 0.78 $\pm$ 0.02 %accuracy 
\\ 

No Primary Training 
& 0.71 $\pm$ 0.04  %b-auroc    
& 0.45 $\pm$ 0.02 %b-auprc
& 0.40 $\pm$ 0.18   %b-precision        
& 0.34 $\pm<$ 0.01  %b-recall
& 0.30 $\pm$ 0.01   %b-F1
& 0.81 $\pm<$ 0.01  %accuracy 
\\

No Adaptive Training  
&  0.65 $\pm$ 0.03  %b-auroc
&  0.43 $\pm$ 0.02 %b-auprc
& 0.41 $\pm$ 0.02  %b-precision
&  0.39 $\pm$ 0.05 %b-recall
&  0.39 $\pm$ 0.03 %b-F1
& 0.66 $\pm$ 0.02  %accuracy 
\\ \cline{2-7}

 \hline        
\end{tabular}\caption{Early-phase ($<3$ day) \textcolor{black}{classification metrics for the observed ZTF BTS Sample, with 0.2-day GP-interpolation scheme, for three cases: no host photometry with the baseline training scheme, host photometry with no primary training, and host photometry with no adaptive training (see text for details). Removing any of these components results in an average decrease in classifier performance across every metric considered.}}\label{tbl:host_and_retrain}
\end{table*}
\end{center}

\begin{table*}[!htp] \hspace*{-12mm}
\begin{tabular}{lllllllll}\hline
Model   & Data & Phase & b-AUROC  & b-AUPRC & b-Precision & b-Recall  & b-F$_1$ Score & Accuracy  \\ \hline \hline
0.2-Day GP 
&  Synthetic 
& \textit{Early}  
& 0.82 $\pm$ 0.01  %b-auroc
& 0.55 $\pm$ 0.02  %b-auprc 
& 0.51 $\pm$ 0.01  %b-precision
& 0.61 $\pm$ 0.03  %b-recall
& 0.54 $\pm$ 0.02  %b-F1 score
& 0.74 $\pm$$<$ 0.01 \\ %accuracy

& 
&  \textit{Intermediate}  
& 0.94 $\pm$ 0.02 %b-auroc
& 0.78 $\pm$ 0.04  %b-auprc - need to change!
& 0.65 $\pm$ 0.02  %b-precision
& 0.78 $\pm$ 0.03  %b-recall
& 0.70 $\pm$ 0.03    %b-F1 score
& 0.85 $\pm$ 0.01\\ %accuracy

&  
& \textit{Late}  
& 0.98 $\pm$ 0.01 %b-auroc
& 0.90 $\pm$ 0.02  %b-auprc 
& 0.77 $\pm$ 0.04  %b-precision
& 0.90 $\pm$ 0.02  %b-recall
& 0.82 $\pm$ 0.03  %b-F1 score
& 0.92 $\pm$ 0.02 \\ \cline{2-9} %accuracy

&  ZTF BTS    
&  \textbf{\textit{Early}} 
&  \textbf{0.74 $\pm$ 0.04}  %b-auroc
&  \textbf{0.52 $\pm$ 0.07}  %b-auprc  
&  \textbf{0.58 $\pm$ 0.13}  %b-precision
&  \textbf{0.46 $\pm$ 0.09}  %b-recall
&  \textbf{0.48 $\pm$ 0.11}  %b-F1 score
& \textbf{0.82 $\pm$ 0.02} \\ %accuracy

& 
& \textit{Intermediate}
& 0.86 $\pm$ 0.06  %b-auroc           
& 0.59 $\pm$ 0.06  %b-auprc 
& 0.65 $\pm$ 0.07  %b-precision  
& 0.58 $\pm$ 0.07  %b-recall
& 0.60 $\pm$ 0.07   %b-F1 score
& 0.84 $\pm$ 0.04 \\  %accuracy

&       
&  \textit{Late}  
&  0.91 $\pm$ 0.04   %b-auroc            
&  0.72 $\pm$ 0.10   %b-auprc 
& 0.72 $\pm$ 0.11    %b-precision
& 0.65 $\pm$ 0.10  %b-recall
&  0.67 $\pm$ 0.10    %b-F1 score
&  0.87 $\pm$ 0.05  \\\hline \hline  %accuracy

0.2-Day GP  
&  ZTF BTS 
&  \textit{Late}  
&  0.91 $\pm$ 0.04   %b-auroc            
&  0.72 $\pm$ 0.10   %b-auprc
& 0.72 $\pm$ 0.11    %b-precision
& 0.65 $\pm$ 0.10  %b-recall
&  0.67 $\pm$ 0.10    %b-F1 score
&  0.87 $\pm$ 0.05  \\  %accuracy

BRF  & ZTF &  \textit{Full}   
& 0.87 $\pm$ 0.02             
& 0.60 $\pm$ 0.05 
& 0.53 $\pm$ 0.03            
& 0.69 $\pm$ 0.05       
& 0.53 $\pm$ 0.04    & -  \\
LSTM RNN  & ZTF &  \textit{Full}  
& 0.88 $\pm$ 0.03            
& 0.65 $\pm$ 0.05 
& 0.57 $\pm$ 0.03            
& 0.68 $\pm$ 0.04            
& 0.58 $\pm$ 0.04     & -\\
TimeModAttn  & ZTF &  \textit{Full}  
& 0.90 $\pm$ 0.02            
& 0.68 $\pm$ 0.06 
& 0.59 $\pm$ 0.02 
& 0.73 $\pm$ 0.04 
& 0.61 $\pm$ 0.03  & -\\ \hline            
\end{tabular}\caption{Classification metrics for the 0.2-Day GP RNN model evaluated on synthetic and observed (ZTF BTS) light curves at early ($t_N<3$ days), intermediate ($3\;\textrm{days}<t_N<15$ days), and late phases ($15\;\textrm{days}<t_N<30$ days). Upper rows indicate the performance of the model in this work, whereas the lower rows indicate only its late-phase performance in comparison with the same metrics at full-phase ($t_N=100$ days) for the BRF, LSTM RNN, and TimeModAttn models from \cite{2022Pimentel_DeepSNClassification}. \textcolor{black}{The bolded row highlights the early performance of the classifier on the observed ZTF sample.} The three models from literature are evaluated on observed ZTF light curves, but the specific SNe considered may vary.}\label{tbl:results}
\end{table*}
We now compare the performance of our classifier to prior methods. We list our balanced performance metrics at each phase bin for the simulated and observed SN samples classified with the 0.2-Day GP model in Table~\ref{tbl:results}.  We acknowledge that few commonly-reported classification metrics are intuitive; for clarity, we also provide the overall accuracy of the network at each phase, although we \textcolor{black}{caution} that this is equal to the micro-averaged precision and so is sensitive to the class imbalance of the dataset ($\sim$80\% SNe~Ia).

In \cite{2022Pimentel_DeepSNClassification}, the authors report the performance of their TimeModAttn models relative to a traditional RNN with LSTM units and a Balanced Random Forest algorithm constructed from 144 features extracted from the light curves. They evaluate these classifiers on a sample of spectroscopically-confirmed SNe from ZTF comparable to that used in this work. We note that these metrics are computed 100 days after detection, 70 days after our longest light curve segment, and so denote them as \textit{Full-Phase}. Using the results from their parallel TimeModAttn network with synthetic pre-training and zero data augmentation, we provide these balanced metrics in Table~\ref{tbl:results}. 

We find no statistically significant difference between the balanced precision, F$_1$ score, AUROC, or AUPRC of our classifier at late-phase ($15<t_N<30$) and those reported for the TimeModAttn model 100 days from trigger, \textcolor{black}{although their mean balanced precision falls below the uncertainty range for ours}. It is likely that our results represent lower limits for the performance of our classifier \textcolor{black}{following 30 days}. We find the TimeModAttn model to be superior to our method only in terms of balanced recall, and we emphasize that this metric is less important than precision for triggering follow-up spectroscopy on targeted events (\textcolor{black}{except in the case of rare, high-value transients}). We find the mean AUROC, AUPRC, precision, and F$_1$ score of our method at late-phase to be superior to the baseline random forest \textcolor{black}{(BRF)} model considered; however, we caution that our statistical uncertainties across samples are larger than each of their considered methods, and this limits our ability to directly compare them. In addition, \cite{2022Pimentel_DeepSNClassification} considered four SN classes (SNe~Ia, SNe~Ib/c, SNe~II, and SLSNe-I), whereas we have excluded  SLSNe-I from our sample. Although the light curve and host galaxy photometry of these events are reasonably distinct from the other classes (and performance is often significantly worse on SNe~Ib/c and SNe~II; these were the two SN classes with the lowest performance across twelve classes considered in \citealt{muthukrishna2019rapid}, \textcolor{black}{while SLSNe achieved the highest late-phase AUROC values in \citealt{2022Pimentel_DeepSNClassification}),} our results may differ when applied to the four-class problem. 

Next, we compare the performance of our network to earlier methods at \textit{partial-phase} classification. The number of classifiers that report early-time performance on real data is low; as a result, we will begin by comparing to the early performance of \texttt{RAPID} and \texttt{SCONE}  on synthetic samples. We caution that both of these classifiers considered more SN classes than this work; in addition, whereas RAPID simulated ZTF light curves, the \texttt{SCONE} framework was tested on LSST-simulated photometry from PLAsTiCC \citep{2019Kessler_PLAstiCC}. The impact to classification of the increased spectral coverage but lower cadence of LSST relative to ZTF is an open question, and should be investigated further.

Figure 7 of \cite{muthukrishna2019rapid} shows the recall of the model across all 12 considered transient classes 2 days since trigger. The performance on SNe~II and SNe~Ib/c is the worst across all classes. Combining these classes into a balanced-recall score, we calculate a value of 0.36, compared to our value of \textcolor{black}{0.61$\ \pm \ $0.03}. Figure 8 presents these values 40 days from trigger, when the model achieves a balanced-recall of 0.57 across SNe~Ia, II, and SNe~Ib/c. Our late-phase \textcolor{black}{balanced recall} on synthetic ZTF samples is \textcolor{black}{0.90$\ \pm\ $0.02}, evaluated at least ten days earlier in phase. The confusion matrices shown in \cite{muthukrishna2019rapid} indicate that Calcium-rich transients and SNe~Iax are significant contaminants for classifying SNe~Ib/c; our superior performance is likely partially attributed to not considering these classes. Nevertheless, we find significantly improved performance on the partial-phase light curves of dominant SN classes relative to RAPID. 

Next, we consider the partial-phase performance of \texttt{SCONE} \citep{2021SCONE} on simulated LSST light curves. Fig.~4 of \cite{2022Scone_PhotometricClassification} indicates that their model achieves (when using redshift) a balanced-recall of 0.75$\ \pm \ $0.03 across SNe~Ia, SNe~II, and SNe~Ib/c 5 days from trigger, compared to our \textcolor{black}{lower} recall of \textcolor{black}{0.61$\ \pm \ $0.03} within the first 3 days. 50 days following trigger, they report a balanced recall of 0.86$\ \pm\ $0.03, compared to our value of \textcolor{black}{0.90$\ \pm\ $0.02} in the first 30 days. Although these results are encouraging, Table~\ref{tbl:results} reveals the degradation in results when transitioning from synthetic to real samples. We strongly encourage the validation of \texttt{RAPID}, \texttt{SCONE}, and other early-time photometric classifiers on spectroscopic SN samples to better understand the advantages of each approach.

In addition to the full-phase metrics reported by \cite{2022Pimentel_DeepSNClassification}, plots are provided for the b-AUROC values of their models as a function of phase. At the earliest observation (approximately three days from trigger), they note a maximum b-AUROC of $\sim50$\% across all of their models on observed ZTF SNe. Our model achieves a significantly higher b-AUROC of $74\pm4$\% within the first 3 days. They report a b-AUROC of $\sim75$\% for their BRF and LSTM RNN models, and $\sim85$\% for their TimeModAttn model at $\sim30$ days from trigger, compared to our \textcolor{black}{$91\pm4$\%}. We conclude that a simple architecture is able to achieve comparable or superior performance to more complex methods\textcolor{black}{, both in distinguishing between SNe~Ia, II, and Ib/c at early phases and in improving performance as more observations are obtained.}

\section{Conclusions \& Future Work}\label{sec:conclusions}
We have shown that an SN classifier leveraging both transient and host-galaxy photometry can achieve comparable performance to machine learning methods with complex architectures. Our algorithm represents one of the first attempts to prepare for real-time SN classification at every stage of the design and implementation pipeline. Below, we outline our primary conclusions: 
\begin{enumerate}
\item Between the three light curve pre-processing methods we considered (`Raw and Pad', `0.2-Day GP', and `100-Timestep GP'), the `0.2-Day GP' model performs \textcolor{black}{marginally better in late-phase classification. The three classifiers achieve comparable early and time-evolving classification when applied to the ZTF BTS data. This suggests that the temporal evolution of the photometry can be equally well-captured by an RNN trained on an interpolated representation of the light curve and the raw photometry itself, if the photometry is high-S/N and suffers from minimal Galactic extinction \textcolor{black}{(as is the case with our high-quality sample)}.}%best in early classification and time-evolving classification when applied to the ZTF BTS. The AUROC and AUPRC metrics increase with phase more clearly in the GP models than the `Raw and Pad' model, which suggests that the temporal evolution of the photometry is better-captured by an RNN when trained on an interpolated representation of the light curve (instead of just learning the temporal evolution from an array of phase values passed in). 

\item Our late-time classification performance (between 15 and 30 days from trigger) is comparable to the leading methods (the balanced random-forest, LSTM RNN, and TimeModAttn networks given in \citealt{2022Pimentel_DeepSNClassification}) 100 days from trigger. The \textcolor{black}{mean} balanced-recall of our method is worse than that of the other methods considered, but this is of secondary importance for prioritizing follow-up of common SN classes for upcoming surveys.
\item Our method, with only a single LSTM layer of 60 gated units, achieves $\sim25$\% higher balanced-AUROC than prior methods within the first three days of observation. \textcolor{black}{Table~\ref{tbl:host_and_retrain} suggests that this is due to a combination of host-galaxy photometry \textit{and} early light curve photometry. The decrease in performance from excluding host-galaxy information was within the uncertainties of the baseline results; other host galaxy properties, such as color, may be more informative.} %Additional work should be done to evaluate whether this early success can be attributed to the color and rise rate of the transient, or the photometry of its host galaxy.
\item The uncertainty in each metric (estimated with 5-fold cross-validation) is higher with our model applied to the ZTF BTS than on synthetic data, and also higher than that reported by \cite{2022Pimentel_DeepSNClassification} for a comparable observational dataset. \textcolor{black}{This may} reflect the variation in the GP model of a single event as more data is collected. If this is indeed the case, it reflects the realistic stochasticity of light curve fitting in real-time. %Nevertheless, this motivates the use of less-flexible physically-motivated light curve models when photometry is scarce.

\item There remains a fundamental mismatch between the performance of  simulations and observations. All three models achieve impressive performance on the synthetic light curves, and \textcolor{black}{suffer a degredation in performance} when applied to real observations. This could be due to overly simplistic host-galaxy correlations and/or overly optimistic redshifts in the simulations, but we predict that classification methods developed with simulated samples (e.g., those for PLAsTiCC and ELAsTiCC) will face an adaptation step to ensure reliable classification on observed samples.
\end{enumerate}

By constructing a normalizing flow to generate redshifts \textcolor{black}{and host-galaxy photometry for every transient in our synthetic sample}, we have increased the realism of the current generation of transient simulations. We have used point estimates for photometric redshifts, but our probabilistic approach provides full posterior distributions for each galaxy. These distributions can reveal bimodalities and other complex behavior not captured by a point estimate, and future work should be devoted to developing classification methods that incorporate this information (e.g., by drawing from the redshift distribution multiple times and quantifying how classification results change \textcolor{black}{with each point estimate, or by passing in a discretized form of the full posterior distribution}).

We have found that classification performance suffered when applying our RNN models directly to the observational data after training with synthetic samples. The performance of our network after re-training suggests that transfer learning can improve the network's robustness to realistically complex datasets. Transfer learning is rapidly becoming a critical component of photometric classification architectures (with \citealt{2022Pimentel_DeepSNClassification} and  \citealt{2022Burhanudin_TransferLearning} released while our network was being developed), and will become increasingly valuable in the initial years of upcoming surveys such as LSST.

Our GP interpolation model is implemented in the lightweight and GPU-accelerated code \texttt{tinyGP}; as a result, it is competitive with the fastest GP models in the literature. When implementing our network for classification directly on the ZTF or LSST alert streams, our architecture can be further optimized to ensure rapid inference. We leave this model refinement to future work.

We have avoided using our GP to interpolate early phases of our transient light curves, except where enough data has been collected to construct a reasonable model for its evolution. The growth of physics-informed machine learning in recent years \citep{2018Wu_PhysicsInformed, 2020Rao_PhysicsInformed, 2022Karpov_PhysicsInformed} represents a promising direction for balancing model flexibility and physically-motivated constraints. Multiple models exist for the photometric evolution of SNe \citep[particularly for SNe~Ia;][]{Kenworthy2021_SALT3, 2022Mandel_bayeSN}, and future work could incorporate one of these models into the mean model of the GP or into a distinct interpolation scheme that is only used at early phases, where limited data is available to condition a model. 

The principles explored in this work have immediate utility for surveys whose science goals include the discovery of young SNe. One of these is the Young SN Experiment \citep[YSE;][]{2021Jones_YSE}, a 1500 deg$^2$ survey that began in 2019 and that uses the Pan-STARRS telescopes to obtain well-sampled $griz$ light curves of low-redshift SNe. A primary goal of this survey is the characterization of early-time flux excesses in young SNe and the construction of the low-redshift SN~Ia anchor sample for upcoming cosmological analyses with the Vera C. Rubin Observatory. Simulations suggest that this survey at full operation will discover $\geq$2 SNe within 3 days of explosion per month, when they can be classified with $\sim 82$\% accuracy using our method. We aim to re-train our classifier using simulated YSE light curves and adapt the network using the light curves consolidated as part of the First Data Release for YSE \textcolor{black}{\citep{2022Aleo_YSEDR1}}. We predict that the increased wavelength coverage and higher cadence achieved by augmenting ZTF light curves with YSE observations will result in superior early classification results to the ones shown in this work. 
%purities of \textcolor{red}{$\sim50\%$ -- update} using our method. The model weights after training on simulated data, as well as the associated code, can be downloaded at \textcolor{red}{put it online} and re-trained on data associated with other low-redshift transient surveys.

Although we have considered only the three dominant classes of SNe in training and testing our network at low-$z$, the problem faced by classifiers operating directly on an alert stream such as that served by LSST will be more difficult: they will have to classify from among the full diversity of observed transient classes spanning $0<z<3$. \textcolor{black}{For this goal, the SCOTCH framework \cite{2022Lokken_SCOTCH} will be a valuable resource. SCOTCH} encodes observed correlations for rarer classes, including Tidal Disruption Events \citep[TDEs;][]{2017French_TDEs, 2020French_TDEs, 2021Zabludoff}, Hydrogen-Poor Superluminous SNe \citep[SLSNe-I;][]{2015Leloudas_SLSN, 2016Perley_SLSNe, 2016Angus_HSTSLSNe}, and Kilonovae \citep[KNe;][]{2006Prochaska_sGRBHosts,2009Berger_sGRBHosts}. These events will represent a fraction of the full stream, but high-cadence coverage of the early-time behavior of even a few events of these classes will be a significant contribution to the literature. Consequently, additional work should be dedicated to extending our classification schema to include these events.

The broad-band photometry of a galaxy in optical pass-bands is, at best, a limited indicator of its underlying spectral energy distribution. Multiple degeneracies exist between, for example, the distance to the galaxy and the color of its average stellar population, particularly across wide ranges in redshift. We have limited our analysis to the approximate redshift range spanned by the ZTF BTS ($z<0.2$) in order to validate our methods on observed data, and it is likely that this has allowed us to largely avoid this drawback \textcolor{black}{(although at these low distances, peculiar velocities introduce additional scatter)}. LSST will dramatically increase our sample of high-redshift ($z<1$) SNe, and at these distances it is unlikely that optical host galaxy photometry alone will have high discriminating power. We have begun investigating the value of host galaxy star-formation rate and stellar mass in simulated samples, but additional work should be devoted to rapid estimation of these parameters from host galaxies imaged with LSST \citep[e.g., using Bayesian SED-fitting codes such as Prospector;][]{2021Johnson_Prospector}. Further, \textcolor{black}{our simulations considered only the photometry of transient hosts in PS1-$grizy$}; for high-redshift systems discovered by the Vera C. Rubin Obs., the \textit{Nancy Grace Roman Space Telescope} \citep{2015Spergel_WFIRST}, and the \textit{James Webb Space Telescope} \citep{2006Gardner_JWST}, infrared photometry will play an increasing role in precise galaxy characterization\textcolor{black}{, as it more precisely traces a galaxy's stellar mass}. Observed transient host galaxy catalogs are increasingly incorporating photometry spanning ultraviolet and infrared wavelengths \citep{2022Qin_THEx}, and future work should be dedicated to reproducing these observations in a simulated host galaxy sample.

\section{Acknowledgements}
We wish to thank G. Narayan, V. Ashley Villar, R. Kessler, J.~F. Crenshaw, A. Rest, and J. Pierel for conversations that improved this work. \textcolor{black}{The authors further thank the anonymous referee for their thorough review, which has significantly strengthened this paper.}

A.G.  acknowledges support from the Flatiron Institute Center for Computational Astrophysics Pre-Doctoral Fellowship Program in Spring 2022. A.G. is also supported by the Illinois Distinguished Fellowship, the National Science Foundation Graduate Research Fellowship Program under Grant No. DGE – 1746047, and the Center for Astrophysical Surveys Graduate Fellowship at the University of Illinois. \textcolor{black}{A.I.M. acknowledges support during this work from the Max Planck Society and the Alexander von Humboldt Foundation in the framework of the Max Planck-Humboldt Research Award endowed by the Federal Ministry of Education and Research. AIM is a member of the LSST Interdisciplinary Network for Collaboration and Computing (LINCC) Frameworks team; LINCC Frameworks is supported by Schmidt Futures, a philanthropic initiative founded by Eric and Wendy Schmidt, as part of the Virtual Institute of Astrophysics (VIA).} P.D.A. is supported by the Center for Astrophysical Surveys at the National Center for Supercomputing Applications (NCSA) as an Illinois Survey Science Graduate Fellow.

This research used resources of the National Energy Research Scientific Computing Center (NERSC), a U.S. Department of Energy Office of Science User Facility located at Lawrence Berkeley National Laboratory, operated under Contract No. DE-AC02-05CH11231.

The ANTARES project has been supported by the National Science Foundation through a cooperative agreement with the Association of Universities for Research in Astronomy (AURA) for the operation of NOIRLab, through an NSF INSPIRE grant to the University of Arizona (CISE AST-1344024, PI: R. Snodgrass), and through a grant from the Heising-Simons Foundation.

ZTF is supported by National Science Foundation grant AST-1440341 and a collaboration including Caltech, IPAC, the Weizmann Institute for Science, the Oskar Klein Center at Stockholm University, the University of Maryland, the University of Washington, Deutsches Elektronen-Synchrotron and Humboldt University, Los Alamos National Laboratories, the TANGO Consortium of Taiwan, the University of Wisconsin at Milwaukee, and Lawrence Berkeley National Laboratories. Operations are conducted by COO, IPAC, and UW. 

The ZTF forced-photometry service was funded under the Heising-Simons Foundation grant \#12540303 (PI: Graham). 

This research made use of the Cori system associated with the National Energy Research Scientific Computing Center (NERSC), a U.S. Department of Energy Office of Science User Facility located at Lawrence Berkeley National Laboratory and operated under Contract No. DE-AC02-05CH11231.% using NERSC award <OOO>-ERCAP<XXXXX>.

\textit{Software:} \texttt{ANTARES} \citep{2021Matheson_ANTARES}, \texttt{Astropy} \citep{astropy:2013, astropy:2018}, \texttt{imbalanced-learn} \citep{JMLR:v18:16-365}, \texttt{Keras} \citep{chollet2015_keras}, \texttt{Matplotlib} \citep{hunter2007matplotlib}, \texttt{numpy} \citep{walt2011_numpy}, \texttt{Pandas} \citep{reback2020_pandas}, \texttt{PZFlow} \citep{2022PZFlow_JFC}, \texttt{Seaborn} \citep{michael_waskom_2014_12710}, \texttt{Scikit-Learn} \citep{scikit-learn}, \texttt{Tensorflow} \citep{tensorflow2015-whitepaper}, \texttt{tinygp}\footnote{\href{https://tinygp.readthedocs.io/en/stable/}{https://tinygp.readthedocs.io/en/stable/}}. 

\bibliography{references}

%\appendix

\end{document}